\documentclass[a4paper,oneside,11pt]{article}

\newtheorem{dfn}{Definition}[section]
\newtheorem{tw}[dfn]{Theorem}
\newtheorem{prop}[dfn]{Proposition}
\newtheorem{rem}[dfn]{Remark}
\newtheorem{ex}[dfn]{Example}
\newtheorem{lem}[dfn]{Lemma}

\usepackage{anysize}
\usepackage{verbatim}
\usepackage{array}
\usepackage{enumerate}
\usepackage{amssymb} 
\usepackage{amsmath}
\usepackage{fancyhdr}
\usepackage{euscript}
\usepackage{latexsym}

\author{Micha\l \ Barski \\ \small  Leipzig University}

\author{Micha\l \ Barski \\ 
\small Faculty of Mathematics and Computer Science, University of Leipzig, Germany\\ 
\small  Faculty of Mathematics, Cardinal Stefan Wyszy\'nski University in Warsaw, Poland\\
\small{\it Michal.Barski@math.uni-leipzig.de} \bigskip \\
}

\title{\bf Monotonicity of the collateralized debt obligations term structure model
}

\begin{document}
\baselineskip=1.1\baselineskip \maketitle
\date
\
\begin{abstract}
The problem of existence of arbitrage free and monotone
CDO term structure models is studied. Conditions for positivity and monotonicity of the corresponding 
Heath-Jarrow-Morton-Musiela equation for the $x$-forward rates with the use of the Milian type 
result are formulated. Two state spaces are taken into account - of square integrable functions and
a Sobolev space. For the first the regularity results concerning pointwise monotonicity  are proven. 
Arbitrage free and monotone models are characterized in 
terms of the volatility of the model and characteristics of the driving L\'evy process.
\end{abstract}

\begin{quote}


\end{quote}


\section{Introduction}
A defaultable $(T,x)$-bond with maturity $T>0$ and credit rating $x\in I\subseteq[0,1]$, a $(T,x)$-bond for short, is a financial contract which pays to its holder $1$ Euro at time $T$ providing that the writer of the bond hasn't bankrupted till time $T$. The set $I$ above stands for all possible credit ratings. The bankruptcy is modeled with the use of a so called loss process $\{L(t), t\geq 0\}$ which starts from zero, increases and takes values in the interval $[0,1]$. The bond is worthless if the loss process exceeds its credit rating. Thus the payoff profile of the $(T,x)$-bond is of the form
$$
\mathbf{1}_{\{L_T\leq x\}}.
$$
The price $P(t,T,x)$ of the $(T,x)$-bond is a stochastic process defined by
\begin{gather}\label{wzor na P}
 P(t,T,x)=\mathbf{1}_{\{L_t\leq x\}}e^{-\int_{t}^{T}f(t,u,x)du}, \quad t\in[0,T],
\end{gather}
where $f(\cdot,\cdot,x)$ stands for an {\it $x$-forward rate}. 
The value $x=1$ corresponds to the risk-free bond and $f(t,T,1)$ determines the {\it short rate} process via
\begin{gather*}
 f(t,t,1), \ t\geq0.
\end{gather*}
The $(T,x)$-bond market is thus fully determined by the family of $x$-forward rates and the loss process $L$. This model is an extension of the classical non-defaultable bond market which can be identified with the case  when $I$ is a singleton, that is, when $I=\{1\}$.

The model of $(T,x)$-bonds above does not correspond to defaultable bonds which are directly traded on a real market. For instance, in this setting the bankruptcy of the $(T,x_2)$-bond automatically implies the bankruptcy of the $(T,x_1)$-bond if $x_1<x_2$. In reality a bond with a higher credit rating may, however, default earlier than that  with a lower one. The $(T,x)$-bonds were introduced  in \cite{FilOverSchm} as basic instruments related to the pool of defaultable assets called Collateralized Debt Obligations (CDO), which are actually widely traded on the market. In the CDO market model the loss process $L(t)$ describes the part of the pool  which has defaulted up to time $t>0$ and $F(L_T)$, where $F$ is some function, specifies the CDO payoff at time $T>0$. In particular, $(T,x)$-bonds may be identified with the digital-type CDO payoffs with the function $F$ of the form
$$
F(z)=F_x(z):=\mathbf{1}_{[0,x]}(z), \qquad x\in I, z\in[0,1].
$$
Then the price of that payoff $p_t(F_x(L_T))$ at time $t\leq T$ equals $P(t,T,x)$. Moreover, as was shown in \cite{FilOverSchm}, each regular CDO claim can be replicated, and thus also priced, with a portfolio consisting of a certain combination of $(T,x)$-bonds. 
Thus it follows that the model of $(T,x)$-bonds determines the structure of the CDO payoffs. The induced  family of prices
$$
P(t,T,x), \qquad T\geq 0, \ x\in I,
$$
will be called a {\it CDO term structure model} or briefly a {\it CDO model}.

On real markets the price of a claim which pays more is always higher. This implies 
\begin{gather}\label{monotonicity in x}
P(t,T,x_1)=p_t(F_{x_1}(L_T))\leq p_t(F_{x_2}(L_T))=P(t,T,x_2), \qquad t\in[0,T], \quad x_1<x_2, \quad x_1,x_2\in I,
\end{gather}
which means that the prices of $(T,x)$-bonds are increasing in $x$. Similarly, if the claim is paid earlier, then it has a higher value and hence 
\begin{gather}\label{monotonicity in T}
P(t,T_1,x)=p_t(F_{x}(L_{T_1}))\geq p_t(F_{x}(L_{T_2}))=P(t,T_2,x), \qquad t\in[0,T_1], \quad T_1<T_2, \quad x\in I,
\end{gather}
which means that the  $(T,x)$-bond prices are decreasing in $T$. The CDO term structure model is called {\it monotone} if 
both conditions \eqref{monotonicity in x}, \eqref{monotonicity in T} are satisfied. Surprisingly, monotonicity of the $(T,x)$-bond prices is not always preserved in mathematical models even if they satisfy severe no-arbitrage conditions, see \cite{FilOverSchm} p.60.  The aim of the paper is to specify the CDO term structure models which are arbitrage-free and monotone. That problem is also studied in \cite{SchmTappe} but in different model settings and with the use of different methods than presented in this paper.

We consider a finite family of $x$-forward rates in the Musiela parametrization  
$$
r(t,z,x_i):=f(t,t+z,x_i), \qquad t,z\geq0, \ x_i\in I, 
$$
with $I=\{0\leq x_1<x_2<...<x_n=1\}$ and study the existence of arbitrage free and monotone CDO models specified by the family of prices
$$
P(t,T,x_i)=\mathbf{1}_{\{L_t\leq x_i\}}e^{-\int_{0}^{T-t}r(t,u,x_i)du}, \qquad t,T\geq 0, \ x_i\in I. 
$$
The forward rate dynamics is given by a stochastic partial differential equation (SPDE) of the form
\begin{gather}\label{HJMM - ogolne}
dr(t)=\big(Ar(t)+F(t,r(t))\big)dt+G(t,r(t-))dZ(t), \quad t\geq 0,
\end{gather}
where $A$ is a differential operator: $Ah(z,x_i)=\frac{\partial}{\partial z}h(z,x_i)$ and $Z$ is a one dimensional L\'evy process.
The drift $F$ is determined by the volatility $G$ and the Laplace exponent of the process $Z$, via the generalized version of the 
Heath-Jarrow-Morton condition. Solutions of the equation \eqref{HJMM - ogolne}, called a Heath-Jarrow-Morton-Musiela equation, are assumed to take values in the Hilbert spaces $\mathbb{L}^{2,\gamma}_n$ or $\mathbb{H}^{1,\gamma}_n$, which means that $r(t,\cdot,x_i)$, for each $x_i\in I$, is a square integrable function, resp. belongs to the Sobolev space of functions with square integrable first derivative. 
From the results in \cite{SchmZab}, see also \cite{FilOverSchm}, one can deduce that existence of an arbitrage free CDO model is equivalent to the solvability of \eqref{HJMM - ogolne} and {\it pointwise monotonicity at zero} of the solution, i.e.
\begin{gather}\label{p. m.}
r(t,0,x_i)\geq r(t,0,x_{i+1}), \quad i=1,2,...,n-1 , \quad \text{for almost all} \ t\geq 0.
\end{gather}
Our approach is based on examining positivity and monotonicity in $x_i\in I$ of the solution to \eqref{HJMM - ogolne}.
Generalizing the result of Milian, see \cite{Milian}, which originally deals with the Wiener process driven SPDEs,  we deduce conditions on the volatility $G$ and jumps of the L\'evy process which are equivalent to positivity and monotonicity of the $\mathbb{L}^{2,\gamma}_n$-valued forward rate solving \eqref{HJMM - ogolne}. These are conditions $(P1),(P2),(M1),(M2)$, see Section \ref{section Formulation of the main results} for a precise formulation, which show that $G$ must satisfy certain growth and Lipschitz-type conditions with constants dependent on possible jumps of the process $Z$. Monotonicity of $r$ in $\mathbb{L}^{2,\gamma}_n$ does not imply \eqref{p. m.}, because $r$ does not have to be pointwise well defined. However, we show that under square integrability condition for $Z$  the solution of \eqref{HJMM - ogolne} actually satisfies \eqref{p. m.} and thus automatically generates an arbitrage free CDO model. Its monotonicity follows from the positivity and monotonicity of the $x$-forward rates. These results are formulated as Theorem \ref{tw dla L^2 glowne} and Proposition \ref{prop point monotonicity}. The conditions providing arbitrage free and monotone CDO models generated by an $\mathbb{H}^{1,\gamma}_n$-valued solution of \eqref{HJMM - ogolne} are formulated in Theorem \ref{tw dla H^1 glowne}. In this case, as we show in Proposition \ref{prop monotonicity of forward rates}, the regularity of elements of $\mathbb{H}^{1,\gamma}_n$ implies that positivity conditions $(P1),(P2)$ are sufficient for the CDO model to be arbitrage free and monotone. We do not need $(M1)$ nor $(M2)$. The results mentioned above need the transformations $F$ and $G$ in \eqref{HJMM - ogolne} to have linear growth and satisfy linear growth conditions. The corresponding conditions in terms of the regularity of $G$ in $\mathbb{L}^{2,\gamma}_n$, resp. $\mathbb{H}^{1,\gamma}_n$ and characteristics of the L\'evy process are formulated in Proposition \ref{prop sufficient dla L^2} and Proposition \ref{prop sufficient dla H^1}.

The paper is structured as follows. In Section \ref{section No arbitrage conditions} we present the preliminary results from \cite{FilOverSchm} and \cite{SchmZab} concerning absence of arbitrage in the CDO model. Here we follow the original papers and use standard parametrization for the $x$-forward rates.  A precise formulation of the monotonicity problem involving the Heath-Jarrow-Morton-Musiela equation is presented in Section \ref{section HJMM equation}. 
Section \ref{section Formulation of the main results} contains formulations of the main results that is Theorem \ref{tw dla L^2 glowne} and Theorem \ref{tw dla H^1 glowne} together with two auxiliary results - Proposition \ref{prop point monotonicity} and
Proposition \ref{prop monotonicity of forward rates} concerning the problem of monotonicity and pointwise monotonicity.
In Subsection \ref{Local Lipschitz conditions and linear growth} we present conditions for linear growth and local Lipschitz conditions which are needed in the main results. Further comments on positivity and monotonicity are presented in Subsection \ref{Further comments on positivity and monotonicity}. Proofs are postponed to Section \ref{Proofs}.

\vskip2mm
\noindent
{{\bf Acknowledgements.} The author would like to thank  A. Rusinek, T. Schmidt, S. Tappe,  and J. Zabczyk for inspiring discussions and helpful suggestions.}

\section{No arbitrage conditions}\label{section No arbitrage conditions}

To explain the model framework of the paper we compile preliminary results from \cite{FilOverSchm} and \cite{SchmZab}. 
They are concerned with the no-arbitrage conditions for the CDO market defined by the forward rates, in a standard parametrization, with the following dynamics 
\begin{gather}\label{dynamics of f}
 df(t,T,x_i)=a(t,T,x_i)dt+b(t,T,x_i)dZ(t),\qquad t>0, \ T>0, \ x_i\in I,
\end{gather}
where $Z$ is a one dimensional L\'evy process and $I=\{x_1,x_2,...,x_n\}$ with $0\leq x_1<x_2<...<x_n=1$.  Equation \eqref{dynamics of f} can be treated as a system of stochastic equations parametrized by maturities $T>0$ and credit ratings $x_i\in I$. The model above was studied in 
\cite{SchmZab} and also in \cite{FilOverSchm} for the case when $Z$ is a Wiener process. In the non-defaultable context, i.e. when $I=\{1\}$ one obtains the classical bond market model setting introduced in \cite{HJM}. 

\noindent
The L\'evy process $Z$ admits the following L\'evy-It\^o decomposition
\begin{align}\label{Levy-Ito decomposition}
Z(t)&=at+qW(t)+ \int_{0}^{t}\int_{\{\mid y\mid\leq 1\}}y\hat{\pi}(ds,dy)+ \int_{0}^{t}\int_{\{\mid y\mid>1\}}y\pi(ds,dy), \quad t\geq0,
\end{align}
where $a\in\mathbb{R}$, $q\geq0$, $W$ is a Wiener process and $(\hat{\pi})$, $\pi$ is a (compensated) Poisson jump measure of $Z$.
Above $\nu$ stands for the L\'evy measure of $Z$, so it satisfies 
$$
\int_\mathbb{R}(\mid y\mid^2 \wedge \ 1)\nu(dy)<+\infty.
$$
The characteristic triplet $(a,q,\nu)$ determines the L\'evy process in a unique way. The central role in the no-arbitrage conditions plays the Laplace transform $J$ of $Z$ which is defined by
\begin{gather}\label{J}
E(e^{-zZ(t)})=e^{tJ(z)},\qquad t\geq 0.
\end{gather}
It is well known that the domain of $J$ is of the form
 $$
 B:=\{z\in\mathbb{R}: \int_{\{\mid y\mid>1\}}e^{-zy}\nu(dy)<+\infty\},
 $$
that is $\mid J(z)\mid<+\infty$ if and only if $z\in B$, see \cite{Sato}, \cite{PeszatZabczyk}. It follows that if $B\neq\emptyset$ then some exponential 
moments of the L\'evy process exist.

\vskip2mm
\noindent

To formulate conditions which are equivalent to the absence of arbitrage on the CDO market, that is which ensure that 
the discounted bond prices 
\begin{gather*}
\hat{P}(t,T,x_i):=e^{-\int_{0}^{t}f(s,s,1)ds}P(t,T,x_i)=e^{-\int_{0}^{t}f(s,s,1)ds}\mathbf{1}_{\{L_t\leq x\}}e^{-\int_{t}^{T}f(t,u,x)du}, \qquad T>0,\ x_i\in I,
\end{gather*}
are local martingales, we need the following set of assumptions (A1)-(A3).
\begin{quote}
{\bf(A1)} \quad  The {\it loss process} $L$ is a c\`adl\`ag, non-decreasing, adapted, pure jump process of the form \ $L_t=\sum_{s\leq t}\triangle L_s, t\geq0$  with absolutely continuous compensator $v(t,dx)dt$ satisfying $\int_{0}^{t}v(s,I)ds<+\infty$.
\end{quote}
Under (A1) the process $\mathbf{1}_{\{L_t\leq x_i\}}$ is c\`adl\`ag for each $x_i\in I$ and has intensity of the form 
$$\lambda(t,x_i):=v(t,(x_i-L_t,1]),$$ 
that is the processes
$$
\mathbf{1}_{\{L_t\leq x_i\}}-\int_{0}^{t}\mathbf{1}_{\{L_s\leq x_i\}}\lambda(s,x_i)ds,
$$
is a martingale. Moreover, $\lambda(t,x_i)$ is progressive and decreasing in $x_i\in I$. 
\begin{quote}
{\bf(A2)} \quad For each $(T,x_i)$ the coefficients $a(t,T,x_i), b(t,T,x_i)$ are predictable and have bounded trajectories.
 \end{quote}
\begin{quote}
{\bf(A3)} \quad For each $r>0$ the function 
 $$
 u\rightarrow \int_{\{\mid y\mid>1\}}e^{-uy}\nu(dy)
 $$
 is bounded on the set $\{u\in\mathbb{R}: \mid u\mid\leq r\}\cap B$. 
  \end{quote}

\noindent
The following result comes from \cite{SchmZab}.

\begin{tw}\label{tw Schmidt-Zabczyk}
 Assume that (A1)-(A3) hold.
 \begin{enumerate}[a)]
  \item If $\hat{P}(t,T,x_i)$, $x_i\in I, T>0$ are local martingales then
  \begin{gather}\label{int b in B}
  \int_{s}^{t}b(s,u,x_i)du\in B
  \end{gather}
  for any $0\leq t\leq s$ on the set $\{L_t\leq x_i\}$, $dP\times dt$ a.s..\\
  \item If \eqref{int b in B} holds then $\hat{P}(t,T,x_i)$, $x_i\in I, T>0$ are local martingales if and only if
\begin{gather}\label{pierwsze}
 \int_{t}^{s}a(t,u,x_i)du=J\left(\int_{t}^{s}b(t,u,x_i)du\right),\\[1ex]\label{drugie}
 f(t,t,x_i)=f(t,t,1)+\lambda(t,x_i),
\end{gather}
for any $0\leq t\leq s$ on the set $\{L_t\leq x_i\}$, $dP\times dt$ a.s..\\
\end{enumerate}
\end{tw}
If $I$ is a singleton, i.e. $I=\{1\}$ and $W$ is a Wiener process then $J(z)=\frac{1}{2}z^2$ and equation \eqref{pierwsze} reduces to the well known Heath-Jarrow-Morton condition from \cite{HJM}. Differentiating \eqref{pierwsze} in $s$ yields the explicit formula for the drift
\begin{gather}\label{pierwsze po modyfikacji}
a(t,T,x_i)=J^\prime\left(\int_{t}^{T}b(t,u,x_i)du\right)b(t,T,x_i),
\end{gather}
in terms of the volatility of the model. Equation \eqref{drugie} reflects the relation between the forward rate  and the distribution of the loss process $L$. It follows that the loss process $L$ may not be given a priori in an arbitrary way. In fact the loss process is uniquely determined by conditions \eqref{pierwsze}, \eqref{drugie}. To see that we directly follow the arguments presented in \cite{FilOverSchm}.
Without loosing generality, we assume that the probability space has the following structure

\begin{gather*}
{\bf (A4)} \quad \Omega=\Omega_1\times\Omega_2, \ \mathcal{F}=\mathcal{G}\otimes\mathcal{H},\ \mathcal{F}_t=\mathcal{G}_t\otimes\mathcal{H}_t, P(d\omega)=P_1(d\omega_1)P_2(\omega_1,d\omega_2), \\ \nonumber 
\text{with} \ \omega=(\omega_1,\omega_2)\in\Omega,  
\end{gather*}
where $(\Omega_1, \mathcal{G}, (\mathcal{G}_t), P_1)$ supports the L\'evy process $Z$ and $\Omega_2$ is the canonical space of increasing, $I$-valued marked point functions endowed with filtration
$$
\mathcal{H}_t:=\sigma\{\omega_2(s): s\leq t, \omega_2\in\Omega_2\}, \ \mathcal{H}:=\mathcal{H}_{\infty}.   
$$
Now one can fix paths of the loss process $\omega_2(t)=L_t(\omega)$ and treat \eqref{dynamics of f} with $a$ satisfying \eqref{pierwsze po modyfikacji} as an equation on 
$(\Omega_1,\mathcal{G}, \mathcal{G}_t, P_1)$. If this equation has a solution, then condition \eqref{drugie} can be written as
$$
v(\omega, t, dx)=-f(\omega,t,t,\omega_2(t)+dx),
$$
which means that the compensator of the loss process is determined by $f$. The problem of determining distribution of the process $L$ is equivalent to finding the probability kernel $P_2(\omega_1,d\omega_2)$ such that $-f(\omega,t,t,\omega_2(t)+dx)$ actually forms a compensator. This holds if $f(t,t,x_i)$ is decreasing in $x_i$. This leads to the following result which is a starting point for further analysis.

\begin{tw}\label{tw ogolne o istnieniu}
Assume that (A1)-(A4) are satisfied.
If \eqref{pierwsze} holds and \eqref{dynamics of f} has a solution for each path of the loss process
$L_t$ such that the function 
\begin{gather}\label{monotonicity of f(t,t,iks)}
x_i\longrightarrow f(t,t,x_i), \quad x_i\in I,
\end{gather}
is decreasing $dP_1\times dt$ a.s. and the process $f(t,t,x_i)$ is progressive then the family $\{f(t,T,x_i); \quad t,T\geq0, \ x_i\in I\}$ forms an arbitrage free CDO model.
\end{tw}

\section{Formulation of the problem}\label{section HJMM equation}
Here we reformulate the dynamics of the $x$-forward rate \eqref{dynamics of f} by passing from the standard paramterization to the Musiela parametrization which was first used in \cite{Musiela}.
 For the running time $t$ and maturity $T$ one defines a new parameter $z=T-t$ called {\it time to maturity}. Then the forward rates  in Musiela parametrization are given by
\begin{gather*}
 r(t,z,x_i):=f(t,t+z,x_i), \qquad t\geq0, z\geq0, \ x_i\in I.
\end{gather*}
and the induced bond prices by
\begin{gather}\label{bond prices Musiela parametrization}
P(t,T,x_i)=\mathbf{1}_{\{L_t\leq x_i\}}e^{-\int_{0}^{T-t}r(t,u,x_i)du},\qquad x_i\in I, \ T\geq0.
\end{gather}
Starting from \eqref{dynamics of f} and using 
\begin{align*}
G(t,z,x_i):=b(t,t+z,x_i), \  F(t,z,x_i):=a(t,t+z,x_i	).
\end{align*}
we obtain
\begin{align}\label{weak solution}
 r(t)(z,x_i)&=S_t(r_0)(z,x_i)+\int_{0}^{t}S_{t-s}F(s)(z,x_i)ds+\int_{0}^{t}S_{t-s}G(s)(z,x_i)dZ(s),
 \end{align}
where $S$ stands for the shift semigroup $S_t(h)(z,x_i):=h(t+z,x_i)$. This means that $r$ is a weak solution 
of the equation
\begin{gather}\label{HJMM z procesem strat general}
 dr(t,z,x_i)=\Big(Ar(t,z,x_i)+F(t,z,x_i)\Big)dt+G(t,z,x_i)dZ(t), \quad t,T\geq 0, x_i\in I,
\end{gather}
with a generator $A$ of the semigroup $S$ given by 
\begin{align*}
 Ar(t,z,x_i):=\frac{\partial r(t,z,x_i)}{\partial z}.
\end{align*} 
The volatility $G$ in \eqref{HJMM z procesem strat general} is assumed to be a transformation of the form $G(t,r(t-))$ with
\begin{gather}\label{volatility}
 G(t,\varphi)(z,x_i)=g(t,z,x_i,L_t,\varphi(z)),
 \qquad t\geq0, \ z\geq0, \  \varphi=\varphi(z),
\end{gather}
where $L_t$ is a loss process and 
\begin{gather}\label{funkcje g}
g(\cdot,\cdot,x_i,\cdot,\cdot)=:g_i(\cdot,\cdot,\cdot,\cdot):\mathbb{R}_+\times\mathbb{R}_+\times I\times\mathbb{R}^n\rightarrow \mathbb{R}, \quad i=1,2,...,n,
\end{gather}
is a sequence of functions. Since we are interested in arbitrage free models only, it follows from \eqref{pierwsze po modyfikacji} that the drift coefficient $F=F(t,r(t))$ in \eqref{HJMM z procesem strat general} is determined by
\begin{gather}\label{dryf HJMM}
 F(t,\varphi)(z,x_i):=J^{\prime}\left(\int_{0}^{z}G(t,\varphi)(u,x_i)du\right)G(t,\varphi)(z,x_i), \qquad t\geq0, \ z\geq0, \varphi=\varphi(z); \ x_i\in I.
\end{gather}

The SPDE \eqref{HJMM z procesem strat general} with volatility $G$ given by \eqref{volatility} and drift $F$ of the form \eqref{dryf HJMM} will be called in the sequel a Heath-Jarrow-Morton-Musiela (HJMM) equation.  It follows that the HJMM equation is specified by $G$ and the function $J^{\prime}$ which in turn is determined by the characteristic triplet of the L\'evy process. 
The HJMM equation in the non-defaultable context has been studied for instance in \cite{BarZab}, \cite{BarZab2}, \cite{FilTappe}, \cite{FilTapTeich}, \cite{Marinelli}. The state space for the solution of \eqref{HJMM z procesem strat general} is to be specified. To this end
let us introduce two Hilbert spaces of measurable real valued functions defined on $\mathbb{R}_{+}$. The first consists of
square integrable functions 
$$
L^{2,\gamma}:=\Big\{h: \ \parallel h\parallel^2_{L^{2,\gamma}}:=\int_{0}^{+\infty}\mid h(u)\mid^2e^{\gamma u}du<+\infty\Big\},
$$
and the second is the Sobolev space - a subspace of  $L^{2,\gamma}$ defined by 
$$
H^{1,\gamma}:=\Big\{h: \ \parallel h\parallel^2_{H^{1,\gamma}}:=\int_{0}^{+\infty}(\mid h(u)\mid^2+\mid h^\prime(u)\mid^2)e^{\gamma u}du<+\infty\Big\},
$$
where $\gamma>0$. The state spaces for the HJMM equation will be $\mathbb{L}^{2,\gamma}_n$ and $\mathbb{H}^{1,\gamma}_n$ consisting of functions
$h:\mathbb{R}_{+}\times I\longrightarrow\mathbb{R}$ such that $h(\cdot,x_i)\in L^{2,\gamma}$, resp. $h(\cdot,x_i)\in H^{1,\gamma}$ for each $x_i\in I$. Endowed with the norms
$$
\parallel h\parallel^2_{\mathbb{L}^{2,\gamma}}:=\sum_{i=1}^{n}\parallel h(\cdot,x_i)\parallel^2_{L^{2,\gamma}}, \quad \parallel h\parallel^2_{\mathbb{H}^{1,\gamma}}:=\sum_{i=1}^{n}\parallel h(\cdot,x_i)\parallel^2_{H^{1,\gamma}}.
$$
they become Hilbert spaces. 

 In view of Theorem \ref{tw ogolne o istnieniu} the CDO model is arbitrage free if there exists a solution of the HJMM equation for each path of the loss process $\{L_t, t\geq0\}$ and such that it is {\it pointwise monotone at zero}, i.e.
\begin{gather}\label{warunek point monotonicity at zero}
r(t,0,x_i)\geq r(t,0,x_{i+1}), \quad i=1,2,...,n-1 , \quad \text{for almost all} \ t\geq 0.
\end{gather}
We additionally require that the $(T,x_i)$-bond prices, given by \eqref{bond prices Musiela parametrization}, are decreasing in $T\geq0$ and increasing in $x_i\in I$.

\section{Formulation of the main results}\label{section Formulation of the main results}

Our conditions which characterize the arbitrage free and monotone CDO term structure models require the transformations 
$G,F$, given by \eqref{volatility} and \eqref{dryf HJMM}, to be locally Lipschitz and to satisfy the linear growth condition (LGC) in $\mathbb{H}$, where $\mathbb{H}$ stands for the state space, i.e. it is equal $\mathbb{L}^{2,\gamma}_n$ or $\mathbb{H}^{1,\gamma}_n$. To be precise, $F,G$ are {\it locally Lipschitz} (LC) if
for any $R>0$ there exists $C_R\geq0$ such that 
\begin{gather}\label{local Lipschitz ogolnie}
\parallel F(t,x)-F(t,y)\parallel_{\mathbb{H}}\leq C_R \parallel x-y\parallel_{\mathbb{H}}, \quad \parallel G(t,x)-G(t,y)\parallel_{\mathbb{H}}\leq C_R \parallel x-y\parallel_{\mathbb{H}}
\end{gather}
for any $ x,y\in\mathbb{H}$ such that $\parallel x\parallel_H, \parallel y\parallel_H\leq R,$ 
and satisfy {\it linear growth condition} if there exists $C\geq0$ such that
\begin{gather}\label{linear growth ogolnie}
\parallel F(t,x)\parallel_{\mathbb{H}}\leq C \parallel x\parallel_{\mathbb{H}}, \quad \parallel G(t,x)\parallel_{\mathbb{H}}\leq C \parallel x\parallel_{\mathbb{H}}
\end{gather}
for any $x,y\in\mathbb{H}$.

\noindent
The first result is concerned with the space $\mathbb{L}^{2,\gamma}_n$. Recall, that $supp\{\nu\}$ stands for the support of the L\'evy measure.
\begin{tw}\label{tw dla L^2 glowne}
Let $(A1)-(A4)$ be satisfied.  Assume that $F$ and $G$ are locally Lipschitz transformations with linear growth in $\mathbb{L}^{2,\gamma}_n$.
Then the following statements hold.
\begin{enumerate}[a)]
\item For any path of the loss process there exists a unique weak solution to the HJMM equation in the space $\mathbb{L}^{2,\gamma}_n$.
\item If for $r=(r_1,r_2,...,r_n)$, $ r_1\geq r_2\geq...\geq r_n, \ t,z\geq0, \ l\in I, \ u\in\emph{supp}\{\nu\}$, $i=1,2,...,n-1$ hold
\begin{align*}
&{\bf(M1)} \qquad &g_i(t,z,l,r)&=g_{i+1}(t,z,l,r), \quad \text{if} \quad r_i=r_{i+1}, \\[1ex]
&{\bf(M2)} \qquad &\Big(g_{i+1}(t,z,l,r)&-g_{i}(t,z,l,r)\Big)u\leq r_i-r_{i+1}.
\end{align*}
and 
\begin{gather}\label{drugi moment nu na zewnetrzu}
\int_{\{\mid y\mid\geq 1\}}\mid y\mid^2\nu(dy)<+\infty,
\end{gather}
then the solution of the HJMM equation is pointwise monotone at zero. Consequently the resulting CDO model is arbitrage free. 
\item If  for $r=(r_1,r_2,...,r_n)$, $r\geq0$, $t,z\geq0, \ l\in I, \ u\in\emph{supp}\{\nu\}$, $i=1,2,...,n$ hold
\vskip-4ex
  \begin{align*}
 &{\bf(P1)} \qquad &g_i(t,z,l,r)&=0, \qquad \text{if} \quad r_i=0,\\[1ex]
 &{\bf(P2)} \qquad &r_i+g_i(t,z,l,r)u&\geq 0.
  \end{align*}
together with $(M1), (M2)$  and \eqref{drugi moment nu na zewnetrzu}, then the resulting CDO model is monotone. 
\end{enumerate}
\end{tw}
\vskip2mm
\noindent
The second result is concerned with the $\mathbb{H}^{1,\gamma}_n$-valued forward rates.
\begin{tw}\label{tw dla H^1 glowne}
Let $(A1)-(A4)$ be satisfied.  Assume that $F$ and $G$ are locally Lipschitz transformations with linear growth in $\mathbb{H}^{1,\gamma}_n$.
Then the following statements hold.
\begin{enumerate}[a)]
\item For any path of the loss process there exists a unique weak solution to the HJMM equation in the space $\mathbb{H}^{1,\gamma}_n$.
\item If $(P1)$ and $(P2)$ hold then the solution of the HJMM equation is pointwise monotone at zero and hence the resulting CDO model is arbitrage free. Moreover, that model is monotone.
\end{enumerate}
\end{tw}

Both points $(a)$ in Theorem \ref{tw dla L^2 glowne} and Theorem \ref{tw dla H^1 glowne} follow directly from the recent result on existence of solution of a general SPDE under locally Lipschitz condition and linear growth, see Theorem 4.1 in \cite{BarZab2}. Section \ref{Local Lipschitz conditions and linear growth} is devoted to the direct specification of the volatility $G$ of the HJMM equation and the characteristic triplet of the L\'evy process for 
\eqref{local Lipschitz ogolnie} and \eqref{linear growth ogolnie} to hold, see Proposition \ref{prop sufficient dla L^2} and Proposition \ref{prop sufficient dla H^1}.

The pairs of conditions $(P1), (P2)$ and $(M1), (M2)$ correspond to positivity and monotonicity of the solution of the HJMM  equation in $\mathbb{L}^{2,\gamma}_n$. They  follow from a generalized version of the result of Milian, see \cite{Milian}, which was concerned with a general SPDE driven by a Wiener process. We show how to pass to a L\'evy process in the case of the HJMM equation. To be more precise, we show in Theorem \ref{tw degeneracy conditions} in Section \ref{Section Monotonicity of forward rates} that $(M1), (M2)$ are equivalent to {\it monotonicity} of $r$, that is for each $t\geq0$
\begin{gather}\label{monotonicity w objasnieniach}
r(t,z,x_i)\geq r(t,z,x_{i+1}), \qquad i=1,2,...,n-1,
\end{gather}
holds for almost all $z\geq 0$, while $(P1), (P2)$ to {\it positivity} of $r$, that is for each $t\geq 0$
\begin{gather}\label{positivity w objasnieniach}
r(t,z,x_i)\geq 0, \quad x_i\in I,
\end{gather}
holds for almost all $z\geq 0$.  A delicate point here is the pointwise monotonicity of the solution at zero required for the CDO model to be arbitrage free. Actually  \eqref{warunek point monotonicity at zero} does not follow from \eqref{monotonicity w objasnieniach}. We call that problem {\it pointwise monotonicity} of the solution in $\mathbb{L}^{2,\gamma}_n$ and solve by proving the following. 
\begin{prop}\label{prop point monotonicity}
Assume that the transformations $F,G: \mathbb{L}^{2,\gamma}_n\rightarrow \mathbb{L}^{2,\gamma}_n$ given by \eqref{dryf HJMM}, \eqref{volatility} are locally Lipschitz and satisfy linear growth condition. Let $Z$ satisfy
\begin{gather}\label{second moment of the measure 2}
\int_{\{\mid y\mid>1\}}\mid y\mid^2 \ \nu(dy)<+\infty.
\end{gather}
 and the solution $r(t), t\geq0$ of \eqref{HJMM z procesem strat general} taking values in $\mathbb{L}^{2,\gamma}_n$ be monotone. Then
 for each $z\geq0$, $i=1,2,...,n-1$ holds
 $$
 r(t,z,x_i)\geq r(t,z,x_{i+1}), \quad \text{for almost all} \ t\geq0.
 $$
\end{prop}
This result clearly implies monotonicity of $r$ at zero and thus statement $(b)$ in Theorem \ref{tw dla L^2 glowne} follows. It is also clear that 
\eqref{monotonicity w objasnieniach} and \eqref{positivity w objasnieniach} imply monotonicity of the bond prices, so $(c)$ in Theorem \ref{tw dla L^2 glowne} holds. Notice that in Theorem \ref{tw dla H^1 glowne} we do not require $(M1)$ nor $(M2)$. Of course, it follows from Theorem \ref{tw dla L^2 glowne} (b) that if $(M1)$ and $(M2)$ hold then the $\mathbb{H}^{1,\gamma}_n$-valued solution is also monotone at zero because $r(t,\cdot,x_i)$ is continuous. From continuity of elements in $\mathbb{H}^{1,\gamma}_n$ follows, however, that
the conditions $(P1)$ and $(P2)$ imply monotonicity of $r$ at zero and monotonicity of the corresponding CDO model at once. More precisely, we prove the following
\begin{prop}\label{prop monotonicity of forward rates}
Let $r(t), t\geq0$ be a positive solution of the HJMM equation in the space $\mathbb{H}^{1,\gamma}_n$.Then the bond prices $P(t,T,x_i)$ are decreasing in $T$, increasing in $x_i$ and $r(t,0,x_i)$ is decreasing in $x_i$ on the set $\{x_i:L_t\leq x_i\}$.
\end{prop}
which implies the assertion $(b)$ of Theorem \ref{tw dla H^1 glowne}.

In Section \ref{Further comments on positivity and monotonicity} we further comment on the conditions $(P1),(P2),(M1),(M1)$ and give an example of a system of functions satisfying them. The detailed presentation dealing with the problem of monotonicity, positivity and pointwise monotonicity of the solution to the HJMM equation is contained in Section  \ref{Section Monotonicity of forward rates}. There we start from the Milian theorem and, by using it to the HJMM equation, show validity of the conditions $(P1),(P2),(M1),(M1)$ in Theorem \ref{tw degeneracy conditions}. Afterwards we prove a sequence of auxiliary results which lead to the proofs of Proposition  \ref{prop point monotonicity} and Proposition \ref{prop monotonicity of forward rates}.

\subsection{Local Lipschitz conditions and linear growth}\label{Local Lipschitz conditions and linear growth}
Here we formulate sufficient conditions for \eqref{local Lipschitz ogolnie} and \eqref{linear growth ogolnie} to hold in 
$\mathbb{L}^{2,\gamma}_n$ and $\mathbb{H}^{1,\gamma}_n$. 
\noindent
For the space $\mathbb{L}^{2,\gamma}_n$ we need the following regularity conditions for $G$

\noindent
{\it\underline{Lipschitz condition}}
\begin{quote}
There exists a constant $C>0$ such that
\begin{gather*}
{\bf(LC)} \qquad \mid g_i(t,z,l,r)-g_i(t,z,l,\bar{r})\mid\leq C \parallel r-\bar{r}\parallel, \qquad
t,z \geq0, \ l\in I, \ r,\bar{r}\in\mathbb{R}^n.
\end{gather*}
\end{quote}

\noindent
{\it \underline{boundedness condition}}
\begin{quote}
There exists $\bar{g}:\mathbb{R}_{+}\longrightarrow\mathbb{R}_{+}$ such that 
\begin{gather*}
{\bf(B1)} \qquad \mid g_i(t,z,l,r)\mid\leq \bar{g}(z), \qquad t,z\geq0, \  x_i,l\in I, \ r\in\mathbb{R}^n,
\end{gather*} 
with $K:=\parallel\bar{g}\parallel_{\mathbb{L}^{2,\gamma}_1}<+\infty$.
\end{quote}

\noindent
{\it \underline{linear growth condition}}
\begin{quote}
There exists a constant $C>0$ such that
\begin{gather*}
{\bf(LGC)} \qquad \mid g_i(t,z,l,r)\mid\leq C\parallel r \parallel, \qquad t,z\geq0, \ 
x_i,l\in I, \ r\in\mathbb{R}^n.
\end{gather*}
\end{quote}
\vskip2mm

\noindent
For the space $\mathbb{L}^{2,\gamma}_n$ we have the following result.
\begin{prop}\label{prop sufficient dla L^2}
Assume that volatility $G$ satisfies  $(LC)$ and
\begin{enumerate}[A)]
\item  $(B1)$ and one of the conditions 
\begin{enumerate}[a)]
 \item $g_i\geq 0, i=1,2,...,n$, \qquad  
$\emph{supp} \{\nu\}\subseteq[-1,+\infty)$ \quad \text{and} \quad $\int_{1}^{+\infty}\mid y\mid^2\nu(dy)<+\infty$,
\item \begin{gather*}
\int_{\{\mid
y\mid\geq1\}}y^2e^{{\frac{K}{\sqrt{\gamma}}}y}\nu(dy)<+\infty,
\end{gather*}
where $K$ is defined in (B1).
\end{enumerate}
\item $(LGC)$, $g_i\geq 0, i=1,2,...,n$ and $Z$ is such that 
\begin{gather}\label{war do lipschitza tw glowne 2}
q=0, \quad \emph{supp} \{\nu\}\subseteq[0,+\infty), \quad \text{and} \quad
\int_{0}^{+\infty}(\mid y\mid\vee\mid y\mid^2)\nu(dy)<+\infty.
\end{gather}
\end{enumerate}
Then $F$ and $G$ satisfy \eqref{local Lipschitz ogolnie} and \eqref{linear growth ogolnie} in $\mathbb{L}^{2,\gamma}_n$.
\end{prop}

For the results in the space $\mathbb{H}^{1,\gamma}_n$ we need further regularity assumptions on $G$, that is, 
more restrictive {\it boundedness} conditions and conditions on the {\it derivatives} of $\{g_i\}$.

\begin{quote}
There exists a constant $C>0$ such that  
\begin{align*}
&{\bf (B2)} \qquad\qquad &&\hat{g}:=\sup_{z}\bar{g}(z)<+\infty,\\[2ex]
&{\bf (B3)} \qquad\qquad &&\mid g(t,z,x_i,l,r)\mid^2\leq C^2\parallel r\parallel,
\end{align*}
where $\bar{g}(z)$ is defined in (B1).
\end{quote}

\begin{quote}
For each $i=1,2,...,n$, the derivatives of $g_i$ satisfy 
\begin{gather*}\label{lipschitzowskosc g_z i g_r}\nonumber
{\bf (D1)} \qquad\qquad\mid g^\prime_z(t,z,x_i,l,r)-g^\prime_z(t,z,x_i,l,\bar{r})\mid+\parallel \triangledown g_i(t,z,x_i,l,r)-\triangledown g_i(t,z,x_i,l,\bar{r})\parallel\\[2ex]
\leq C\parallel r-\bar{r}\parallel,
\end{gather*}
\begin{gather*}\label{ograniczenie g_z}
{\bf (D2)} \qquad\qquad \mid g^\prime_z(t,z,x_i,l,r)\mid\leq h(z), \qquad t,z\geq0,  \ x_i,l\in I, r\in\mathbb{R},
\end{gather*} 
for some $h:\mathbb{R}_{+}\longrightarrow \mathbb{R}_{+}$ 
such that $h\in \mathbb{L}^{2,\gamma}_1$ and
\begin{gather*}\label{ograniczenie g_z przez stala}
{\bf (D3)} \qquad\qquad\sup_{z}\mid h(z)\mid<C,
\end{gather*}
and
\begin{gather*}\label{ograniczenie g_r}
{\bf (D4)} \qquad\qquad\sup_{t,z,l,r}\parallel \triangledown g_i(t,z,x_i,l,r)\parallel<C,
\end{gather*}
where
\begin{displaymath}
\triangledown g_i(t,z,l,r):=\left(
\begin{array}{l}
\frac{d}{dr_1}\ g_i(t,z,l,r)\\[1ex]
\frac{d}{dr_2}\ g_i(t,z,l,r)\\[1ex]
...\\
...\\[1ex]
\frac{d}{dr_n}\ g_i(t,z,l,r)
\end{array}\right), \qquad i=1,2,...,n.
\end{displaymath}
\end{quote}

\begin{prop}\label{prop sufficient dla H^1}
Assume that volatility $G$ satisfies $(LC)$, $(D1)-(D4)$ and
\begin{enumerate}[A)]
\item$(B1)-(B3)$ and one of the conditions
\begin{enumerate}[a)]
\item  $g_i\geq 0$, $i=1,2,...,n$,  \qquad $\emph{supp} \{\nu\}\subseteq[-1,+\infty)$  \ \text{and} \ $\int_{1}^{+\infty}\mid y\mid^3\nu(dy)<+\infty$, 
 \item \begin{gather*}
\int_{\{\mid
y\mid\geq1\}}y^3e^{{\frac{K}{\sqrt{\gamma}}}y}\nu(dy)<+\infty,
\end{gather*}
\end{enumerate}
where $K$ is defined in (B1).
\item $(LGC)$,
$(B3)$ and $g_i\geq 0, i=1,2,...,n$ together with 
\begin{gather*}
 \sup_{t,z,l,r}\mid g_i(t,z,l,r)\mid <+\infty, \quad i=1,2,...,n.
\end{gather*}
Further, let  $Z$ be such that 
\begin{gather*}
q=0, \quad \emph{supp} \{\nu\}\subseteq[0,+\infty), \quad \text{and} \quad
\int_{0}^{+\infty}(\mid y\mid\vee\mid y\mid^3)\nu(dy)<+\infty.
\end{gather*}
\end{enumerate}
Then $F$ and $G$ satisfy \eqref{local Lipschitz ogolnie} and \eqref{linear growth ogolnie} in $\mathbb{H}^{1,\gamma}_n$.
\end{prop}

\vskip2mm
\noindent
Let us comment the results above.

\begin{rem}
The L\'evy process satisfying condition $(B)$ in Proposition \ref{prop sufficient dla L^2} or $(B)$ in Proposition \ref{prop sufficient dla H^1} is a subordinator with drift, the L\'evy measure of which additionally satisfies $\int_{\{y>1\}}y^2\nu(dy)<+\infty$, resp. $\int_{\{y>1\}}y^3\nu(dy)<+\infty$.
\end{rem}

\begin{rem}
If the jumps of the L\'evy process $Z$ are bounded then all the assumptions concerning the L\'evy measure in Proposition \ref{prop sufficient dla L^2} and Proposition \ref{prop sufficient dla H^1} are satisfied.
\end{rem}

\noindent
The proofs of Proposition \ref{prop sufficient dla L^2} and Proposition \ref{prop sufficient dla H^1} are postponed to Section \ref{section Local Lipschitz conditions and linear growth proofs}.

\subsection{Further comments on positivity and monotonicity}\label{Further comments on positivity and monotonicity}

Let us start with an observation concerning the case when the HJMM equation split into a separable system of equations.

\begin{rem} Assume that each function $g_i$ in \eqref{HJMM z procesem strat general}, as a function of $r\in\mathbb{R}^n$, depends on the $i$-th coordinate of $r$ only, that is 
$$
g_i(t,z,l,r)=g_i(t,z,l,r_i), \qquad t,z\geq0, \ l\in I, i=1,2,...,n.
$$
Then $(M1)$ holds if and only if the system $\{g_i\}_i$ reduces to one function, that is
$$
g_i(t,z,l,r)=g_j(t,z,l,r), \qquad t,z\geq0, \ l\in I,
$$
for each $i,j=1,2,...,n$. This means that only a trivial system preserves monotonicity of forward rates.
\end{rem}
\vskip2mm
\noindent
Now we provide an auxiliary result dealing with conditions (P1), (P2), (M1), (M2). To abbreviate the notation set $\mathbf{1}_{i}(r):=(r_1,r_2,...,r_{i-1},0,r_{i+1},...,r_n)$ for $r\in\mathbb{R}^n$. 
\begin{prop}\label{prop o konkretnych warunkach na positivity}
  \begin{enumerate}[A)]
  \item  Assume that $g_i\geq0,i=1,2,...,n$.
  \begin{enumerate}[a)]
   \item If (P1) and (P2) hold, then
   \begin{gather}\label{war konieczny positivity - g rozniczkowalna}
  \emph{supp}\{\nu\}\subseteq\left[-\frac{1}{\sup_{t,z,l,r}g^{\prime}_{r_i}(t,z,l,x_i,\mathbf{1}_{i}(r))},+\infty\right),
\end{gather}
for $t,z\geq0$, $l\in I$, $r=(r_1,r_2,...,r_n)\geq0$ and $i=1,2,...,n$.
\item If (P1), \eqref{war konieczny positivity - g rozniczkowalna} hold and
\begin{gather}\label{war dostateczny positivity - g rozniczkowalna}
 g_i(t,z,l,r)\leq g^{\prime}_{r_i}(t,z,x_i,l,\mathbf{1}_{i}(r))r_i, 
\end{gather}
with $g^{\prime}_{r_i}(t,z,x_i,l,\mathbf{1}_{i}(r))\geq0$ for $t,z\geq0$, $l\in I$, $r=(r_1,r_2,...,r_n)\geq0$ $i=1,2,...,n$, then (P2) holds.
 \end{enumerate}
 \item
 \begin{enumerate}[a)]
  \item If (M1), (M2) hold, then for $r=(r_1,r_2,...,r_n), r_1\geq r_2\geq...\geq r_n, \ t,z\geq0, \ l\in I, \ u\in\emph{supp}\{\nu\}$, $i=1,2,...,n-1$, hold
  \begin{gather}\label{pierwszy war na roznice pochodnych}
  \frac{d}{dr_i}[g_{i+1}(t,z,l,r)-g_i(t,z,l,r)]u\leq 1, \quad \text{for} \quad r_i=r_{i+1},
  \end{gather}
  and
  \begin{gather}\label{drugi war na roznice pochodnych}
  \frac{d}{dr_{i+1}}[g_{i+1}(t,z,l,r)-g_i(t,z,l,r)]u\leq 1, \quad \text{for} \quad r_i=r_{i+1},
  \end{gather}
  \item Assume that (M1) and \eqref{pierwszy war na roznice pochodnych} hold. If, for $r=(r_1,r_2,...,r_n), r_1\geq r_2\geq...\geq r_n, \ t,z\geq0, \ l\in I, \ u\in\emph{supp}\{\nu\}$, $i=1,2,...,n-1$, one of the following conditions is satisfied
  \begin{enumerate}[(i)]
  \item  $g_{i+1}(t,z,l,r)-g_i(t,z,l,r)$ is concave in $r_i$ and $\text{supp}\{\nu\}\subseteq(0,+\infty)$,
  \item $g_{i+1}(t,z,l,r)-g_i(t,z,l,r)$ is convex in $r_{i}$  and $\text{supp}\{\nu\}\subseteq(-\infty,0)$, 
  \item $g_{i+1}(t,z,l,r)-g_i(t,z,l,r)$ is concave in $r_i$  and $g_{i+1}(t,z,l,r)\geq g_i(t,z,l,r)$,
  \item $g_{i+1}(t,z,l,r)-g_i(t,z,l,r)$ is convex in $r_i$  and $g_{i+1}(t,z,l,r)\leq g_i(t,z,l,r)$,
  \end{enumerate}
  then (M2) holds.
   \end{enumerate}
   \end{enumerate}
\end{prop}
\vskip2mm
\noindent
With the use of Proposition \ref{prop o konkretnych warunkach na positivity} we can construct the following example.
\noindent
\begin{ex}\label{przyklad}
Let us consider a system of functions of the multiplicative form
$$
g_i(t,z,l,r):=f_1(t)f_2(z)f_3(l)h_1(r_1)h_2(r_2)...h_n(r_n)h(r_i); \qquad i=1,2,...,n.
$$
with smooth functions $f_i, h_i, h$ and the following conditions
\begin{gather}\label{P1}
f_i,h_i,h\geq0, \qquad  f_i\leq\bar{f}_i, h_i\leq\bar{h}_i, \qquad \text{where} \ \bar{f}_i,\bar{h}_i\in\mathbb{R}_{+} \quad i=1,2,...,n,
\end{gather}
\begin{gather}\label{P2}
 h(0)=0, \qquad h^{\prime}(0)\geq0, \qquad h(r_i)\leq h^\prime(0)r_i, \ r_i\geq0,
\end{gather}
\begin{gather}\label{P3}
h_i \ \text{is decreasing} \ i=1,2,...,n,
\end{gather}
\begin{gather}\label{P4}
supp\{\nu\}\subseteq \Big[-\frac{1}{a},+\infty\Big), \quad \text{with} \ a:=\max_i\bar{f}_1\bar{f}_2\bar{f}_3\bar{h}_1...\bar{h}_{i-1}h_i(0)\bar{h}_{i+1}...\bar{h}_nh^\prime(0),
\end{gather}
\begin{gather}\label{P5}
0\leq h^\prime\leq \bar{h^\prime}, \quad \text{where} \ \bar{h^\prime}\in\mathbb{R}_{+}, \qquad \text{and} \quad h,h_i \ \text {are concave for} \ i=1,2,...,n,
\end{gather}
\begin{gather}\label{P6}
supp\{\nu\}\subseteq \Big[-\frac{1}{b},+\infty\Big), \ \text{where} \ b:=\bar{f}_1\bar{f}_2\bar{f}_3\bar{h}_1...\bar{h}_n\bar{h^\prime}.
\end{gather}
Then
\begin{enumerate}[a)]
 \item (P1) and (P2) hold if \eqref{P1}-\eqref{P4} are satisfied.
 \item (P1), (P2), (M1) and (M2) hold if \eqref{P1}-\eqref{P6} are satisfied.
\end{enumerate}
\end{ex}
\vskip2mm
\noindent
The proof of Proposition \ref{prop o konkretnych warunkach na positivity} and calculations concerning Example \ref{przyklad} are postponed to  Section \ref{section ostatni}.

\section{Proofs}\label{Proofs}

\subsection{Local Lipschitz conditions and linear growth}\label{section Local Lipschitz conditions and linear growth proofs}
Here we will prove Proposition \ref{prop sufficient dla L^2} and Proposition \ref{prop sufficient dla H^1}. Let us start with the properties of the Laplace transform defined in \eqref{J}. It is well known that $J$ can be represented in the form 
\begin{gather}\label{Laplace transform}
J(z)=-az+\frac{1}{2}qz^2+\int_{\mathbb{R}}(e^{-zy}-1+zy\mathbf{1}_{(-1,
1)}(y)) \ \nu(dy),\qquad z\in B,
\end{gather}
see \cite{Sato}, \cite{PeszatZabczyk}.  Moreover, the first and second derivative of $J$ exist providing that 
corresponding exponential moment exist, see for instance \cite{Rusinek}. In the sequel we will use the following result which can be proven directly.

\begin{lem}\label{lemat 3}
 The functions $J^\prime$, $J^{\prime\prime}$ are
 \begin{enumerate}[a)]
  \item continuous on $[0,+\infty)$ if
  $$
  \emph{supp} \{\nu\}\subseteq[-1,+\infty), \quad \text{and} \quad
\int_{1}^{+\infty}\mid y\mid^2\nu(dy)<+\infty.
  $$
  \item continuous on $[-z_0,z_0]$ for some $z_0>0$ if
  $$
  \int_{\mid
y\mid\geq1}\mid y\mid ^2e^{z_0\mid y\mid}\nu(dy)<+\infty.
$$
\item continuous and bounded on $[0,+\infty)$ if $Z$ does not contain the Wiener part, i.e. $q=0$, and
 $$
  \emph{supp} \{\nu\}\subseteq[0,+\infty), \quad \text{and} \quad
\int_{0}^{+\infty}(\mid y\mid\vee\mid y\mid^2)\nu(dy)<+\infty.
  $$
 \end{enumerate}
\end{lem}


First let us prove Proposition \ref{prop sufficient dla L^2}. For the sake of notational convenience all the estimations are presented in the equivalent coordinate form, that is for the transformations
$$
\mathbb{L}^{2,\gamma}_n\ni\varphi(\cdot) \rightarrow G(t,\varphi)(\cdot, x_i)\in L^{2,\gamma}, \qquad
\mathbb{L}^{2,\gamma}_n\ni\varphi(\cdot) \rightarrow F(t,\varphi)(\cdot, x_i)\in L^{2,\gamma}, \qquad i=1,2,...,n.
$$

 If (B1) holds, then for any
$\varphi:\mathbb{R}\longrightarrow\mathbb{R}^n$ the following inequality holds
\begin{align}\label{estimation for the integral}\nonumber
&\int_{0}^{z}\mid g_i(t,v,L_t,\varphi(v))\mid dv\leq
\int_{0}^{z}\bar{g}(v)dv\leq
\int_{0}^{+\infty}e^{-\frac{\gamma}{2}v}e^{\frac{\gamma}{2}v}\bar{g}(v)dv\\[2ex]
&\leq\sqrt{\int_{0}^{+\infty}e^{-\gamma
v}dv}\sqrt{\int_{0}^{+\infty}e^{\gamma v}\bar{g}^2(v)dv}\leq
\frac{K}{\sqrt{\gamma}}.
\end{align}

\noindent
{\bf Proof of Proposition \ref{prop sufficient dla L^2}:} 
$(A)$ \quad First we show linear growth. For any $\varphi\in \mathbb{L}^{2,\gamma}_n$ we have
\begin{gather*}
\|G(t,\varphi)(x_i)\|^2_{{L}^{2,\gamma}}=\int_{0}^{+\infty}g^2_i(t,z,l,\varphi(z))e^{\gamma
z}dz\leq\int_{0}^{+\infty}\bar{g}^2(z)e^{\gamma
z}dz=\|\bar{g}\|^2_{L^{2,\gamma}}.
\end{gather*}
\noindent
$(Aa)$ \quad The assumption $g\geq0$ allows us to consider the function
$J^{\prime}$ restricted to $[0,+\infty)$. It follows from condition (Aa) and
Lemma \ref{lemat 3} that $J^{\prime}$ is well defined and increasing on
$[0,+\infty)$. Thus by \eqref{estimation for the integral}
we have
\begin{align*}
\|F(t,\varphi)(x_i)\|^2_{L^{2,\gamma}}&=\int_{0}^{+\infty}\Big|J^{\prime}\left(\int_{0}^{z}g_i(t,v,L_t,\varphi(v))dv\right)g_i(t,z,L_t,\varphi(z))\Big|^2e^{\gamma
z}dz\\[2ex]
&\leq
\Big|J^{\prime}\left(\frac{K}{\sqrt{\gamma}}\right)\Big|^2\int_{0}^{+\infty}\Big|\bar{g}(z)\Big|^2e^{\gamma
z}dz=\Big|J^{\prime}\left({\frac{K}{\sqrt{\gamma}}}\right)\Big|^2\|\bar{g}\|^2_{L^{2,\gamma}}.
\end{align*}

\noindent
Let $\varphi,\phi\in \mathbb{L}^{2,\gamma}_n$.  In view of
$(LC)$ we obtain
\begin{align}\label{lipschitz for g}\nonumber
\|G(t,\varphi)(x_i)-G(t,\phi)(x_i)\|^2_{{L}^{2,\gamma}}&=\int_{0}^{+\infty}\Big|g_i(t,z,L_t,\varphi(z))-g_i(t,z,L_t,\phi(z))\Big|^2e^{\gamma
z}dz\\[2ex]&\leq
C^2\int_{0}^{+\infty}\big\|\varphi(z)-\phi(z)\big\|^2e^{\gamma
z}dz=C^2\|\varphi-\phi\|^2_{\mathbb{L}^{2,\gamma}_n},
\end{align}
and
\begin{align*}
&\|F(t,\varphi)(x_i)-F(t,\phi)(x_i)\|^2_{{L}^{2,\gamma}}\\[1ex]&=\|J^{\prime}\left(\int_{0}^{z}g_i(t,v,L_t,\varphi(v))dv\right)g_i(t,z,L_t,\varphi(z))-
J^{\prime}\left(\int_{0}^{z}g_i(t,v,L_t,\phi(v))dv\right)g_i(t,z,L_t,\phi(z))\|^2_{{L}^{2,\gamma}}\\[2ex]
&\leq 2I_1(x_i)+2I_2(x_i),
\end{align*}
where
\begin{align}\label{definicja I1}
 I_1(x_i)&:=\Big\|J^{\prime}\left(\int_{0}^{z}g_i(t,v,L_t,\varphi(v))dv\right)\Big|g_i(t,z,L_t,\varphi(z))-g_i(t,z,L_t,\phi(z))\Big|\Big\|^2_{{L}^{2,\gamma}},\\[1ex]\label{definicja I2}
 I_2(x_i)&:=\Big\|g_i(t,z,L_t,\phi(z))\Big|J^{\prime}\left(\int_{0}^{z}g_i(t,v,L_t,\varphi(v))dv\right)-J^{\prime}\left(\int_{0}^{z}g_i(t,v,L_t,\phi(v))dv\right)\Big|\Big\|^2_{{L}^{2,\gamma}}.
\end{align}
It follows from Lemma \ref{lemat 3} that under (Aa) the function $J^{\prime\prime}$ is well defined and
continuous on $[0,+\infty)$. Thus $J^{\prime}$ is Lipschitz on any
interval $[0,a]$, $a>0$. Let $C(J^{\prime},{\frac{K}{\sqrt{\gamma}}})$
denote the Lipschitz constant of the function $J^{\prime}$ on the
interval $[0,{\frac{K}{\sqrt{\gamma}}}]$. By \eqref{estimation for the integral} and
(LC) we have 
\begin{align*}
 I_1(x_i)\leq \Big|J^{\prime}\left({\frac{K}{\sqrt{\gamma}}}\right)\Big|^2
C^2\int_{0}^{+\infty}\parallel\varphi(z)-\phi(z)\parallel^2e^{\gamma
z}dz=\Big|J^{\prime}\left({\frac{K}{\sqrt{\gamma}}}\right)\Big|^2
C^2\|\varphi-\phi\|^2_{\mathbb{L}^{2,\gamma}_n},
\end{align*}
\begin{align*}
 I_2(x_i)&\leq \Big\|g_i(t,z,L_t,\phi(z))C\left(J^{\prime},{\frac{K}{\sqrt{\gamma}}}\right)\int_{0}^{+\infty}\mid
g_i(t,v,L_t,\varphi(v))-g_i(t,v,L_t,\phi(v))\mid
dv\Big\|^2_{{L}^{2,\gamma}}\\[1ex]
&\leq\Big\|g_i(t,z,L_t,\phi(z))C\left(J^{\prime},{\frac{K}{\sqrt{\gamma}}}\right)C\int_{0}^{+\infty}\parallel
\varphi(v)-\phi(v)\parallel dv\Big\|^2_{{L}^{2,\gamma}}\\[1ex]
&\leq
C^2\left(J^{\prime},{\frac{K}{\sqrt{\gamma}}}\right)\frac{C^2}{\gamma}\|\bar{g}\|^2_{{L}^{2,\gamma}}\|\varphi-\phi\|^2_{\mathbb{L}^{2,\gamma}_n},
\end{align*}
and the assertion follows.\\
\noindent
$(Ab)$
If we drop the positivity assumption of  $\{g_i\}$, then the
proof above can be mimicked, but, in view of \eqref{estimation for
the integral}, we have to know that the function $J^{\prime}$ is
well defined and Lipschitz on
$[-{\frac{K}{\sqrt{\gamma}}},{\frac{K}
{\sqrt{\gamma}}}]$ for each $i=1,2,...,n$. In view of
Lemma \ref{lemat 3} we need to assume $(Ab)$.
\vskip2mm
\noindent
$(B)$ \quad By $(LGC)$ for any $\varphi\in \mathbb{L}^{2,\gamma}_n$ holds
\begin{gather*}
\|G(t,\varphi)(x_i)\|^2_{{L}^{2,\gamma}}=\int_{0}^{+\infty}g_i^2(t,z,L_t,\varphi(z))e^{\gamma
z}dz\leq
C^2\int_{0}^{+\infty}\parallel\varphi(z)\parallel^2 e^{\gamma
z}dz=C^2\|\varphi\|^2_{\mathbb{L}^{2,\gamma}_n}.
\end{gather*}
The assumption $g\geq0$ allows us to consider the functions
$J^{\prime}, J^{\prime\prime}$ restricted to $[0,+\infty)$. It follows from Lemma
\ref{lemat 3} that if $Z$ has no Wiener part and \eqref{war do lipschitza tw glowne 2} holds, then
$J^{\prime}, J^{\prime\prime}$ are well defined on $[0,+\infty)$ and bounded, i.e.
$$
\sup_{z\geq0}\mid J^{\prime}(z)\mid\leq M, \qquad \sup_{z\geq0}\mid J^{\prime\prime}(z)\mid\leq M,
$$
for some $M>0$. Thus, in view of (LGC), we have

\begin{align*}
\|F(t,\varphi)(x_i)\|^2_{{L}^{2,\gamma}}&=\int_{0}^{+\infty}\Big|J^{\prime}\left(\int_{0}^{z}g_i(t,v,L_t,\varphi(v))dv\right)g_i(t,z,L_t,\varphi(z))\Big|^2e^{\gamma
z}dz\\[2ex]
&\leq M^2C^2\int_{0}^{+\infty}\parallel\varphi(z)\parallel^2e^{\gamma
z}dz=M^2C^2\|\varphi\|^2_{\mathbb{L}^{2,\gamma}_n}.
\end{align*}

\noindent
Let $\varphi,\phi\in \mathbb{L}^{2,\gamma}_n$.  In view of
$(LC)$ we obtain
\begin{align}\label{lipschitz for g}\nonumber
\|G(t,\varphi)(x_i)-G(t,\phi)(x_i)\|^2_{{L}^{2,\gamma}}&=\int_{0}^{+\infty}\Big|g_i(t,z,L_t,\varphi(z))-g_i(t,z,L_t,\phi(z))\Big|^2e^{\gamma
z}dz\\[2ex]&\leq
C^2\int_{0}^{+\infty}\parallel\varphi(z)-\phi(z)\parallel^2e^{\gamma
z}dz=C^2\|\varphi-\phi\|^2_{\mathbb{L}^{2,\gamma}_n}.
\end{align}
To show that $F$ satisfies local Lipschitz condition, first let us notice
that, in view of (LGC), for any $\|\varphi\|_{\mathbb{L}^{2,\gamma}_n}\leq R$ the following estimate holds
\begin{gather}\nonumber
 \int_{0}^{z}\mid g_i(t,v,L_t,\varphi(z))\mid dv\leq C\int_{0}^{z}\parallel \varphi(v)\parallel dv\\[1ex]\label{aaa}
 \leq C\sqrt{\int_{0}^{+\infty}e^{-\gamma v}dv}\sqrt{\int_{0}^{+\infty}e^{\gamma v}\parallel\varphi(v)\parallel^2dv}\leq \frac{CR}{\sqrt{\gamma}}, \quad z>0.
\end{gather}

\noindent
We have
\begin{gather*}
\|F(t,\varphi)(x_i)-F(t,\phi)(x_i)\|^2_{{L}^{2,\gamma}}\leq 2I_1(x_i)+2I_2(x_i),
\end{gather*}
where $I_1(x_i),I_2(x_i)$ are defined in \eqref{definicja I1},
\eqref{definicja I2}. Using $(LC)$, $(LGC)$ and \eqref{aaa} we obtain
\begin{align*}
I_1(x_i)&\leq
M^2 C^2\int_{0}^{+\infty}\parallel\varphi(z)-\phi(z)\parallel^2e^{\gamma
z}dz=M^2 C^2\|\varphi-\phi\|^2_{\mathbb{L}^{2,\gamma}_n},
\end{align*}
and
\begin{align*}
I_2(x_i)&\leq\Big\|g_i(t,z,L_t,\phi(z))M\int_{0}^{+\infty}\mid
g_i(t,v,L_t,\varphi(v))-g_i(t,v,L_t,\phi(v))\mid
dv\Big\|^2_{{L}^{2,\gamma}}\\[2ex]
&\leq
M^2C^2\left\{\int_{0}^{+\infty}\parallel
\varphi(v)-\phi(v)\parallel dv\right\}^2C^2\|\phi\|^2_{\mathbb{L}^{2,\gamma}_n}\\[2ex]
&\leq M^2\frac{C^4}{\gamma}\|\varphi-\phi\|^2_{\mathbb{L}^{2,\gamma}_n}\|\phi\|^2_{\mathbb{L}^{2,\gamma}_n}\leq M^2\frac{C^4}{\gamma}R^2\|\varphi-\phi\|^2_{\mathbb{L}^{2,\gamma}_n},
\end{align*}
and thus local Lipschitz condition for $F$ follows.  \hfill$\square$

\vskip2mm
Now we pass to the proof of Proposition \ref{prop sufficient dla H^1}. We examine the transformations 
$$
\mathbb{H}^{1,\gamma}_n\ni\varphi(\cdot) \rightarrow G(t,\varphi)(\cdot, x_i)\in H^{1,\gamma}, \qquad
\mathbb{H}^{1,\gamma}_n\ni\varphi(\cdot) \rightarrow F(t,\varphi)(\cdot, x_i)\in H^{1,\gamma}, \qquad i=1,2,...,n,
$$
and use the estimations from the proof  of Proposition \ref{prop sufficient dla L^2}.  Assume that $F$ and $G$ satisfy Lipschitz condition in $\mathbb{L}^{2,\gamma}_n$. Then it follows from the formula
$$
\|h\|^2_{H^{1,\gamma}}=\int_{0}^{+\infty}\Big(h^2(z)+(h^\prime(z))^2\Big)e^{\gamma z}dz=\|h\|^2_{L^{2,\gamma}}+\|h^{\prime}\|^2_{L^{2,\gamma}},
$$
that
\begin{align*}
\parallel G(t,\varphi)(\cdot,x_i)&-G(t,\phi)(\cdot,x_i)\parallel^2_{H^{1,\gamma}}\\[2ex]
&=\parallel G(t,\varphi)(\cdot,x_i)-G(t,\phi)(\cdot,x_i)\parallel^2_{L^{2,\gamma}}
+\Big\| \frac{d}{dz}G(t,\varphi)(\cdot,x_i)-\frac{d}{dz}G(t,\phi)(\cdot,x_i)\Big\|^2_{L^{2,\gamma}}\\[2ex]
&\leq C\parallel\varphi-\phi\parallel^2_{\mathbb{L}^{2,\gamma}_n}+\Big\| \frac{d}{dz}G(t,\varphi)(\cdot,x_i)-\frac{d}{dz}G(t,\phi)(\cdot,x_i)\Big\|^2_{L^{2,\gamma}}\\[2ex]
&\leq C\parallel\varphi-\phi\parallel^2_{\mathbb{H}^{1,\gamma}_n}+\Big\| \frac{d}{dz}G(t,\varphi)(\cdot,x_i)-\frac{d}{dz}G(t,\phi)(\cdot,x_i)\Big\|^2_{L^{2,\gamma}}.
\end{align*}
Thus to get Lipschitz conditions in $\mathbb{H}^{1,\gamma}_n$ we will study transformations
$$
\mathbb{H}^{1,\gamma}_n\ni\varphi\longrightarrow \frac{d}{dz}G(t,\varphi)(\cdot,x_i)\in L^{2,\gamma}, \qquad \mathbb{H}^{1,\gamma}_n\ni\varphi\longrightarrow \frac{d}{dz}F(t,\varphi)(\cdot,x_i)\in L^{2,\gamma},
$$
which, in view of \eqref{volatility} and \eqref{dryf HJMM}, are given by
\begin{align}\label{G prime}
\frac{d}{dz}G(t,\varphi)(z,x_i)&=g^{\prime}_z(t,z,x_i,L_t,\varphi(z))+\langle \triangledown g_i(t,z,L_t,\varphi(z)),\varphi^{\prime}(z)\rangle,\\[2ex]\label{F prime}\nonumber
\frac{d}{dz}F(t,\varphi)(z,x_i)&=J^{\prime\prime}\left(\int_{0}^{z}g_i(t,u,L_t,\varphi(u))du\right)g_i^2(t,z,L_t,\varphi(z))\\[1ex]
&+J^{\prime}\left(\int_{0}^{z}g_i(t,u,L_t,\varphi(u))du\right)\Big[g^{\prime}_z(t,z,x_i,L_t,\varphi(z))+\langle \triangledown g_i(t,z,L_t,\varphi(z)),\varphi^{\prime}(z)\rangle\Big].
\end{align}
Above $\langle \cdot,\cdot\rangle$ stands for the scalar product in $\mathbb{R}^n$.

Let us start with an auxiliary inequality
\begin{gather}\label{auxilliary lemma}
\sup_{z\geq0}\|\varphi(z)\|\leq \frac{2}{\sqrt{\gamma}}\|\varphi\|_{\mathbb{H}^{1,\gamma}_n}, \qquad \varphi\in\mathbb{H}^{1,\gamma}_n,
\end{gather}
which follows from the inequality
$$
\sup_{z\geq0}\mid \varphi_i(z)\mid\leq \frac{2}{\sqrt{\gamma}}\|\varphi_i\|_{H^{1,\gamma}},
$$
proved in \cite{BarZab2}, see Lemma 4.4.

\vskip2mm

\noindent
{\bf Proof of Proposition \ref{prop sufficient dla H^1}:} $(A)$ \quad First we show that (D2) and (D4) imply linear growth of $\frac{d}{dz}G$.
\begin{align*}
\|\frac{d}{dz}G(t,\varphi)(x_i)\|^2_{{L}^{2,\gamma}}&=\int_{0}^{+\infty}[g^\prime_z(t,z,x_i,L_t,\varphi(z))+\langle \triangledown g_i(t,z,L_t,\varphi(z)),\varphi^\prime(z)\rangle]^2e^{\gamma z}dz\\[2ex]
&\leq 2\int_{0}^{+\infty}\mid h(z)\mid^2e^{\gamma z}dz+2\sup_{t,z,l,r}\parallel \triangledown g_i(t,z,x_i,l,r)\parallel^2\int_{0}^{+\infty}\parallel \varphi^\prime(z)\parallel^2e^{\gamma z}dz\\[2ex]
&\leq 2\|h\|^2_{{L}^{2,\gamma}}+2C^2\|\varphi\|^2_{\mathbb{H}^{1,\gamma}_n}.
\end{align*}
\noindent
To show linear growth of  $\frac{d}{dz}F$ recall that the assumptions in (a) or in (b) imply that 
$$
J^{\prime\prime}\Big(\int_{0}^{z}g_i(t,v,L_t,\varphi(u))dv\Big), \qquad J^{\prime}\Big(\int_{0}^{z}g_i(t,v,L_t,\varphi(u))dv\Big),
$$
are bounded on $\mathbb{R}$, so in view of the formula \eqref{F prime} we additionally need to show that $g^2_i$ has linear growth. 
By (B3) we have 
$$
\int_{0}^{+\infty}\mid g_i^2(t,z,L_t,\varphi(z))\mid^2e^{\gamma z}dz \leq C^4\int_{0}^{+\infty}\parallel\varphi(z)\parallel^2e^{\gamma z}dz\leq C^4\|\varphi\|^2_{\mathbb{H}^{1,\gamma}_n},
$$
and linear growth of $F$ follows.
 Now we will  prove Lipschitz estimates. In view of (D1), (D4) and \eqref{auxilliary lemma} we have
 \begin{align*}
 &\left\|\frac{d}{dz}\Big(G(t,\varphi)(x_i)-G(t,\phi)(x_i)\Big)\right\|^2_{L^{2,\gamma}}\leq 2\int_{0}^{+\infty}[g^\prime_z(t,z,x_i,L_t,\varphi(z))-g^\prime_z(t,z,x_i,L_t,\phi(z))]^2e^{\gamma z}dz\\[2ex]
 &\qquad+2\int_{0}^{+\infty}[\langle \triangledown g_i(t,z,L_t,\varphi(z)),\varphi^\prime(z)\rangle-\langle \triangledown g_i(t,z,L_t,\phi(z)),\phi^\prime(z)\rangle]^2e^{\gamma z}dz\\[2ex]
 &\qquad\leq 2C^2\|\varphi-\phi\|^2_{\mathbb{H}^{1,\gamma}_n}+4\int_{0}^{+\infty}\Big[\langle \triangledown g_i(t,z,L_t,\varphi(z)),\varphi^\prime(z)-\phi^\prime(z)\rangle\Big]^2e^{\gamma z}dz\\[2ex]
&\qquad+4\int_{0}^{+\infty}\Big[\langle \phi^\prime(z), \triangledown g_i(t,z,L_t,\varphi(z))-\triangledown g_i(t,z,L_t,\phi(z))\rangle\Big]^2e^{\gamma z}dz\\[2ex]
 \end{align*}
 \begin{align*}
&\qquad\leq 2C^2\|\varphi-\phi\|^2_{\mathbb{H}^{1,\gamma}_n}+4C^2\|\varphi-\phi\|^2_{\mathbb{H}^{1,\gamma}_n}\\[2ex]
&\qquad+4\sup_{t,z,l,r,\bar{r}}\frac{\mid g^{\prime}_{r}(t,z,x_i,r)-g^{\prime}_{r}(t,z,x_i,\bar{r})\mid^2}{\parallel r-\bar{r}\parallel^2}\int_{0}^{+\infty}\parallel\phi^\prime(z)\parallel^2\cdot
\parallel\varphi(z)-\phi(z)\parallel^2e^{\gamma z}dz\\[2ex]
&\qquad\leq\left(6C^2+\frac{16}{\gamma}C^2\|\phi\|^2_{\mathbb{H}^{1,\gamma}_n}\right)\|\varphi-\phi\|^2_{\mathbb{H}^{1,\gamma}_n},
 \end{align*}
and local Lipschitz property of $\frac{d}{dz}G$ follows.
It follows from \eqref{F prime} that
\begin{gather*}
\left\|\frac{d}{dz}\Big(F(t,\varphi)(x_i)-F(t,\phi)(x_i)\Big)\right\|^2_{L^{2,\gamma}}\leq 3(I_1(x_i)+I_2(x_i)+I_3(x_i)),
\end{gather*}
where
{\footnotesize
\begin{align*}
I_1(x_i)&:=\left\|J^{\prime\prime}\left(\int_{0}^{z}g_i(t,u,L_t,\varphi(u))du\right)g^2_i(t,z,L_t,\varphi(z))-
J^{\prime\prime}\left(\int_{0}^{z}g_i(t,u,L_t,\phi(u))du\right)g^2_i(t,z,L_t,\phi(z))\right\|^2_{L^{2,\gamma}}\\[1ex]
I_2(x_i)&:=\left\|J^{\prime}\left(\int_{0}^{z}g_i(t,u,L_t,\varphi(u))du\right)g^\prime_z(t,z,x_i,L_t,\varphi(z))-
J^{\prime}\left(\int_{0}^{z}g_i(t,u,L_t,\phi(u))du\right)g^\prime_z(t,z,x_i,L_t,\phi(z))\right\|^2_{L^{2,\gamma}}\\[1ex]
I_3(x_i)&:=\left\|J^{\prime}\left(\int_{0}^{z}g_i(t,u,L_t,\varphi(u))du\right)\langle\triangledown g_i(t,z,L_t,\varphi(z)),\varphi^\prime(z)\rangle-
J^{\prime}\left(\int_{0}^{z}g_i(t,u,L_t,\phi(u))du\right)\langle\triangledown g_i(t,z,L_t,\phi(z)),\phi^\prime(z)\rangle\right\|^2_{L^{2,\gamma}}
\end{align*}}
We have
\begin{align*}
I_1(x_i)&\leq 2\left\|\mid J^{\prime\prime}\left(\int_{0}^{z}g_i(t,u,L_t,\varphi(u))du\right)\mid\cdot\mid g^2_i(t,z,L_t,\varphi(z))-g^2_i(t,z,L_t,\phi(z))\mid\right\|^2_{L^{2,\gamma}}\\[1ex]
&+2\left\|g^2_i(t,z,L_t,\phi(z))\mid J^{\prime\prime}\left(\int_{0}^{z}g_i(t,u,L_t,\varphi(u))du\right)-J^{\prime\prime}\left(\int_{0}^{z}g_i(t,u,L_t,\phi(u))du\right)\mid\right\|^2_{L^{2,\gamma}}.
\end{align*}
It follows from (LC), (B1) and (B2) that 
\begin{gather*}
\mid g^2_i(t,z,L_t,r)-g^2_i(t,z,L_t,\bar{r})\mid\leq 2C\hat{g}\parallel r-\bar{r}\parallel
\end{gather*}
and thus the first expression above can be estimated. For the second we use again (B2) and the fact that the assumptions in (Aa) or (Ab) imply that $ J^{\prime\prime\prime}$ is locally bounded on a positive half-line or around zero, respectively. The estimate for $I_2(x_i)$ follows from  (D1) and (D3) while (D1) and (D4) imply local Lipschitz property for $I_3(x_i)$. 
\vskip2mm
\noindent
$(B)$ The proof is similar to part $(A)$. The only difference is that the assumptions on the L\'evy measure ensure that $J^{\prime},J^{\prime\prime},J^{\prime\prime\prime}$ are bounded on $[0,+\infty)$.
\hfill$\square$

\subsection{Monotonicity of the forward rates}\label{Section Monotonicity of forward rates}
In this section we present the results on positivity, monotonicity and pointwise monotonicity of the solution to \eqref{HJMM z procesem strat general}. Our final aim is to prove Proposition \ref{prop point monotonicity} and Proposition \ref{prop monotonicity of forward rates}.

We start with an auxiliary result on positivity and monotonicity of a general system of equations of the form

\begin{align}\label{rownanie Milian1}\nonumber
 dX_1&=(AX_1+F_1(X_1,X_2,...,X_n))dt+G_1(X_1,X_2,...,X_n)dW,\\
 dX_2&=(AX_2+F_2(X_1,X_2,...,X_n))dt+G_2(X_1,X_2,...,X_n)dW,\\ \nonumber
 &\vdots\\ \nonumber
 dX_n&=(AX_n+F_n(X_1,X_2,...,X_n))dt+G_n(X_1,X_2,...,X_n)dW,\nonumber
 \end{align}
where $W$ is a one dimensional Wiener process and
$$
F_i,G_i:H^n\rightarrow H, \qquad i=1,2,...,n,
$$
with the space $H$ of square integrable functions on some measurable space $E$ with a sigma-finite measure.
The solution to \eqref{rownanie Milian1} is assumed to be an element of $H^n$. The following is a version of the Milian result, see \cite{Milian}. 

$C(E)$, $C_c^{\infty}(E)$ below stand for the space of continuous functions and smooth functions with compact support respectively.

\begin{tw}[Milian]\label{tw Milian}
 Assume that $A$ generates a strongly continuous semigroup $S_t, t\geq0$ in $H$ and that the semigroup preserves positivity.
Assume that for each $R>0$ there exists a constant $C_R$ such that
  \begin{gather}\label{Lipschitz conditions in Milian theorem}
   \|F_i(x)-F_i(y)\|_{H}+\|G_i(x)-G_i(y)\|_{H}\leq C_R\|x-y\|_{H^n},\quad i=1,2,...,n,
     \end{gather}
  for each $x,y\in B_R:=\{z\in H^n: \parallel z\parallel_{H^n}\leq R\}$. Assume that there exists a solution $X$ to \eqref{rownanie Milian1}.

\begin{enumerate}[a)]
\item Let $X(0)\geq0$. If for each $f\in H_{+}\cap C_{c}^{\infty}(E)$ and $\varphi=(\varphi_1,...,\varphi_n)$, $\varphi_i\in
H_{+}\cap C(E), i=1,2,...,n$ such that $\langle \varphi_i,f\rangle=0$ for some $i=1,2,...,n$ the following
holds
\begin{gather}\label{Milian condition 1a}
 \langle F_i(\varphi),f\rangle\geq 0,\\ \label{Milian condition 1b}
\langle G_i(\varphi),f\rangle=0,
\end{gather}
then $X(t)\geq0, t\geq0$. Moreover, if $X(t)\geq 0, t\geq0$ then \eqref{Milian condition
1a} and \eqref{Milian condition 1b} hold.

\item Let $X_1(0)\geq X_2(0)\geq...\geq X_n(0)$. If for each $f\in H_{+}\cap C_{c}^{\infty}(E)$ and $\varphi=(\varphi_1,...,\varphi_n)$, $\varphi_i\in H\cap C(E), i=1,2,...,n$ such that
$\varphi_1\geq\varphi_2\geq,...,\geq\varphi_n$ and 
$\langle\varphi_{i},f\rangle=\langle\varphi_{i+1},f\rangle$ for some $i=1,2,...,n-1$ the following holds
\begin{gather}\label{Milian condition 2a}
 \langle F_i(\varphi),f\rangle\geq \langle F_{i+1}(\varphi),f\rangle\\ \label{Milian condition 2b}
 \langle G_i(\varphi),f \rangle=\langle G_{i+1}(\varphi),f\rangle,
\end{gather}
then $X_i(t)\geq X_{i+1}(t),t\geq0, i=1,2,...,n-1$. Moreover, if $X_i(t)\geq X_{i+1}(t),t\geq0, i=1,2,...,n-1$ then
\eqref{Milian condition 2a} and \eqref{Milian condition 2b} hold.
\end{enumerate}
\end{tw}

In the original formulation condition \eqref{Lipschitz conditions in Milian theorem} is replaced by a global Lipschitz condition. Conditions for positivity of a general SPDE under locally Lipschitz conditions were proven in \cite{BarZab2}, see Theorem 4.2. Monotonicity under locally Lipschitz conditions can be shown in a similar way. 

Using Theorem \ref{tw Milian} we characterize monotonicity and positivity of solutions to the HJMM equation with L\'evy noise.

\begin{tw}\label{tw degeneracy conditions} Assume that the transformations $G,F$ given by \eqref{volatility} and \eqref{dryf HJMM} are locally Lipschitz in $\mathbb{L}^{2,\gamma}_{n}$ and  let $r(t), t\geq0$, be a solution to the HJMM equation  in the space $\mathbb{L}^{2,\gamma}_n$. The following statements hold.

 \begin{enumerate}[a)]
 \item If $r_0\geq0$, then $r(t)\geq 0, t\geq0$ if and only if both conditions $(P1)$ and $(P2)$ are satisfied.
 \item Assume that $r_0(z,x_i)$ is decreasing in $x_i\in I$. Then $r(t,z,x_i)$ is decreasing in $x_i\in I$ if and only if both conditions $(M1)$ and $(M2)$ are satisfied.
 \end{enumerate}
\end{tw}

\noindent
{\bf Proof of Theorem \ref{tw degeneracy conditions}:} The solution to \eqref{HJMM z procesem strat general} is positive and monotone if and only if for each $\varepsilon\in(0,1)$ is the solution $r^{\varepsilon}(t)$ of the system
\begin{align}\label{HJMM epsilon}\nonumber
 dr^\varepsilon(t,z,x_i)&=
 \Big(Ar^\varepsilon(t,z,x_i)+F(t,r^\varepsilon)(z,x_i)+(a-m_\varepsilon)G(t,r^\varepsilon)(z,x_i)\Big)dt\\[1ex]
 &+qG(t,r^\varepsilon)(z,x_i)dW(t)+G(t,r^\varepsilon)(z,x_i)dP^{\varepsilon}(t),
\end{align}
with
$$
m_\varepsilon:=\int_{\{\varepsilon<y<1\}}y\nu(dy), \qquad P^\varepsilon(t)=\int_{0}^{t}\int_{\{y>\varepsilon\}}y\pi(ds,dy).
$$
Equation \eqref{HJMM epsilon} arises from the original equation \eqref{HJMM z procesem strat general} by cutting out compensated
jumps smaller than $\varepsilon$. Since $P^\varepsilon$ is a compound Poisson process, it has a finite number of jumps on each finite time interval and between the jumps the driving noise is the Wiener process. Thus to get positivity and monotonicity one can directly use Theorem \ref{tw Milian}. 

\noindent
$(a)$ First we show necessity of (P1). Fix any $x_i\in I$.  Condition
\eqref{Milian condition 1b} applied with
$\varphi=(\varphi_1,...,\varphi_n)$, s.t. $\varphi_i\equiv 0$ provides, that for any $f\in H_{+}\cap
C^{\infty}_{c}(\mathbb{R}_{+})$ we
have
\begin{gather*}
\int_{0}^{+\infty}g_i(t,z,L_t,\varphi(z))f(z)e^{\gamma z}dz=0,
\end{gather*}
which implies $g_i(t,z,L_t,\varphi(z))=0, z\geq0$. This gives (P1).

To show sufficiency of (P1) we will check conditions \eqref{Milian condition 1a} and \eqref{Milian condition 1b}. Let $f\in H_{+}\cap C^{\infty}_{c}(\mathbb{R}_{+})$ and $\varphi=(\varphi_1,...,\varphi_n)$, $\varphi\in H_{+}\cap C(\mathbb{R}_{+})$ be such that
\begin{gather*}
 \int_{0}^{+\infty}\varphi_i(z)f(z)e^{\gamma z}dz=0,
\end{gather*}
for some $i=1,2,...,n.$
Then
\begin{gather*}
 \lambda(A_i\cap B)=0, \quad \text{where} \quad A_i:=\{z: \varphi_i(z)>0\}, \ B:=\{z: f(z)>0\},
\end{gather*}
and $\lambda$ stands for the Lebesgue measure. Using
(P1) we have
\begin{align*}
 \langle F(t,\varphi)(x_i)&+(a-m_\varepsilon)G(t,\varphi)(x_i),f\rangle\\[1ex
 ]&=\int_{0}^{+\infty}J^{\prime}\left(\left(\int_{0}^{z}g_i(t,v,L_t,\varphi(v))dv\right)+a-m_\varepsilon\right)g_i(t,z,L_t,\varphi(z))f(z)e^{\gamma
 z}dz\\[1ex]
&=\int_{A_i}J^{\prime}\left(\left(\int_{0}^{z}g_i(t,v,L_t,\varphi(v))dv\right)+a-m_\varepsilon\right)g_i(t,z,L_t,\varphi(z))f(z)e^{\gamma
 z}dz\\[1ex]
 &+\int_{B}J^{\prime}\left(\left(\int_{0}^{z}g_i(t,v,L_t,\varphi(v))dv\right)+a-m_\varepsilon\right)g_i(t,z,L_t,\varphi(z))f(z)e^{\gamma
 z}dz=0,
\end{align*}
because both integrals are equal to zero. Moreover, it holds
\begin{align*}
 \mid\langle G(t,\varphi)(x_i),f\rangle\mid&\leq \int_{0}^{+\infty}\mid g_i(t,z,L_t,\varphi(z))\mid f(z)e^{\gamma
 z}dz\\[1ex]
 &\leq\int_{A_i}\mid g_i(t,z,L_t,\varphi(z))\mid f(z)e^{\gamma z}dz+\int_{B}\mid g_i(t,z,L_t,\varphi(z))\mid f(z)e^{\gamma
 z}dz=0.
\end{align*}
At the moment of jump of $P^\varepsilon$ the solution remains positive if 
\begin{gather*}
r_i+g_i(t,z,L_t,r)u\geq0, \quad t,z\geq0, r\in\mathbb{R}^n, r\geq0,  \ u\in \text{supp}\{\nu\}\cap (\varepsilon,+\infty), i=1,2,...,n.
\end{gather*}
Passing to the limit $\varepsilon\downarrow0$ we obtain (P2).

\vskip2mm
\noindent $(b)$
To show necessity of (M1) let us examine
condition \eqref{Milian condition 2b}. Then for 
$\varphi=(\varphi_1,...,\varphi_n)$ with $\varphi_i\in H\cap C(\mathbb{R}_{+}), \forall i$ such that $\varphi_i=\varphi_{i+1}$ for some $i$ and any $f\in
H_{+}\cap C_{c}^{\infty}(\mathbb{R}_{+})$ holds
\begin{gather*}
\int_{0}^{+\infty}\Big(g_{i+1}(t,z,L_t,\varphi(z))-g_{i}(t,z,L_t,\varphi(z))\Big)f(z)e^{\gamma
z}dz=0,
\end{gather*}
which is equivalent to (M1). To show sufficiency of (M1), consider $\varphi=(\varphi_1,...,\varphi_n), \varphi_i\in H\cap C(\mathbb{R}_{+}), \forall i$, $f\in H_{+}\cap C^{\infty}_{c}(\mathbb{R}_{+})$ such that $\varphi_i\geq\varphi_{i+1}, \forall i$ and
\begin{gather*}
 \int_{0}^{+\infty}\Big(\varphi_i(z)-\varphi_{i+1}(z)\Big)f(z)e^{\gamma z}dz=0, \qquad \text{for some} \ i=1,...,n.
\end{gather*}
Then
\begin{gather*}
 \lambda(B\cap\{z:\varphi_i(z)>\varphi_{i+1}(z)\})=0, \quad \text{where} \ B:=\{z: f(z)>0\},
\end{gather*}
and thus we have
\begin{align*}
 \langle F(t,\varphi)(x_{i+1})&+(a-m_\varepsilon)G(t,\varphi)(x_{i+1}),f\rangle\\[1ex]
 &=\int_{0}^{+\infty}J^{\prime}\left(\left(\int_{0}^{z}g_{i+1}(t,v,L_t,\varphi(v))dv\right)+a-m_\varepsilon\right)g_{i+1}(t,z,L_t,\varphi(z))f(z)e^{\gamma
 z}dz\\[1ex]
&=\int_{B}J^{\prime}\left(\left(\int_{0}^{z}g_{i+1}(t,v,L_t,\varphi(v))dv\right)+a-m_\varepsilon\right)g_{i+1}(t,z,L_t,\varphi(z))f(z)e^{\gamma
 z}dz\\[1ex]
&=\int_{0}^{+\infty}J^{\prime}\left(\left(\int_{0}^{z}g_i(t,v,L_t,\varphi(v))dv\right)+a-m_\varepsilon\right)g_i(t,z,L_t,\varphi(z))f(z)e^{\gamma
 z}dz\\[1ex]
&=\langle F(t,\varphi)(x_i)+(a-m_\varepsilon)G(t,\varphi)(x_{i}),f\rangle
\end{align*}
and
\begin{align*}
\langle G(t,\varphi)(x_{i+1}),f \rangle &=\int_{B}g_{i+1}(t,z,L_t,\varphi(z))f(z)e^{\gamma z}dz=\int_{B}g_i(t,z,L_t,\varphi(z))f(z)e^{\gamma z}dz\\[1ex]
 &=\langle G(t,\varphi)(x_i),f\rangle.
\end{align*}
Hence conditions \eqref{Milian condition 2a} and \eqref{Milian condition 2b} are satisfied.
At the moments of jumps the solution remains monotone if for each consecutive pair of coordinates $i,i+1$ holds
\begin{gather*}
r_i+g_{i}(t,z,L_t,r) u\geq r_{i+1}+g_{i+1}(t,z,L_t,r)u,\\[1ex]
t,z\geq 0,  \ r\in\mathbb{R}^n, \ r_i\geq r_{i+1}, \forall i, \ u\in\text{supp}\{\nu\}\cap (\varepsilon,+\infty).
\end{gather*}
Letting $\varepsilon\downarrow 0$  yields (M2).\hfill$\square$

\vskip2mm
Now we pass to the pointwise monotonicity of the solution to the HJMM equation taking values in $\mathbb{L}^{2,\gamma}_n$.
Here we adopt the idea presented in \cite{Rusinek2} and instead of studying irregular functions of $z$
$$
z\rightarrow r(t,z,x_i),
$$
which are defined only for almost each $z$, we consider the functions
$$
z\rightarrow \int_{0}^{t}r(s,z,x_i)ds,
$$
which we prove to be well defined for each $z\geq0$ and regular.  The following Proposition \ref{prop jednoznacznosc calki w L2} and Proposition \ref{prop positivity in L2} lead to to the proof of Proposition \ref{prop point monotonicity}.

\begin{prop}\label{prop jednoznacznosc calki w L2} 
Let $r(t), t\geq0$ be a solution of the HJMM equation in $\mathbb{L}^{2,\gamma}_n$ with coefficients $F$ and $G$ which are locally Lipschitz and have linear growth. Assume that $Z$ is a square integrable L\'evy process, that is
\begin{gather}\label{second moment of the measure}
\int_{\{\mid y\mid>1\}}\mid y\mid^2 \ \nu(dy)<+\infty.
\end{gather}
Then for each $z\geq0$, $x_i\in I$ and $t\geq0$ the function
 $$
 z\rightarrow \int_{0}^{t}r(s,z,x_i)ds,
 $$
 is well defined. Moreover, for each $z\geq0$ and a sequence $z_n\underset{n}{\longrightarrow} z$ there exists a subsequence $z_{n_k}, k=1,2,...$ such that 
 \begin{gather}\label{regularity of the integral}
 \int_{0}^{t}r(s,z_{n_k},x_i)ds\underset{k}{\longrightarrow} \int_{0}^{t}r(s,z,x_i)ds.
 \end{gather}
\end{prop}
{\bf Proof:} For the sake of brevity we use the notation $F_i(t,r(t))(z):=F(t,r(t))(z,x_i)$, $G_i(t,r(t-))(z):=G(t,r(t-))(z,x_i)$. Integrating both sides of \eqref{weak solution} and using the form of the semigroup $S$ we obtain
\begin{align}\label{postac slabego rownania w dowodzie 2}\nonumber
 \int_{0}^{t}r(s,z,x_i)ds=\int_{0}^{t}r(0,z+s,x_i)ds&+\int_{0}^{t}\int_{0}^{s}F_i(u,r(u))(z+s-u)duds\\[1ex]
 &+\int_{0}^{t}\int_{0}^{s}G_i(u,r(u-))(z+s-u)dZ(u)ds.
\end{align}
Now we will argue that the Fubini theorem and the stochastic Fubini theorem can be applied.
We have
\begin{align*}
\int_{0}^{t}\int_{z}^{z+t-u}&\mid F_i(u,r(u))(v)\mid dvdu\leq \int_{0}^{t}\int_{0}^{+\infty}\mid F_i(u,r(u))(v)\mid dvdu\\[1ex]
&\leq\frac{1}{\sqrt{\gamma}}\int_{0}^{t}\|F_i(u,r(u))\|_{L^{2,\gamma}}du\leq\frac{C}{\sqrt{\gamma}}\int_{0}^{t}(1+\|r(u)\|_{\mathbb{L}^{2,\gamma}_n})du<+\infty,
\end{align*}
where the last inequality follows from the fact that $r(t), t\geq 0$ is c\`adl\`ag in $\mathbb{L}^{2,\gamma}_n$ and thus also bounded on bounded intervals. Further, we have
\begin{align*}
E\int_{0}^{t}\int_{z}^{z+t-u}&\mid G_i(u,r(u-))(v)\mid^2 dvdu\leq E\int_{0}^{t}\int_{0}^{+\infty}\mid G_i(u,r(u-))(v)\mid^2 dvdu\\[1ex]
&\leq E\int_{0}^{t}\|G_i(u,r(u-))\|^2_{L^{2,\gamma}}du\leq C^2(x_i)\int_{0}^{t}E(1+\|r(u-)\|_{\mathbb{L}^{2,\gamma}_n})^2du<+\infty,
\end{align*}
where the last inequality follows from the fact $\sup_{u\in[0,t]}E\|r(u)\|^2_{\mathbb{L}^{2,\gamma}_n}<+\infty$, see Theorem 9.29 in \cite{PeszatZabczyk}. Applying the deterministic and stochastic Fubini theorems in \eqref{postac slabego rownania w dowodzie 2}, 
see Theorem 8.14 in \cite{PeszatZabczyk}, we obtain
\begin{align}\label{rozklad calki w twierdzeniu}\nonumber
\int_{0}^{t}r(s,z,x_i)ds=\int_{z}^{z+t}r(0,s,x_i)ds&+\int_{0}^{t}\int_{z}^{z+t-u}F_i(u,r(u))(v)dvdu\\[1ex]
&+\int_{0}^{t}\int_{z}^{z+t-u}G_i(u,r(u-))(v)dvdZ(u).
\end{align}
Since the right hand side is uniquely defined for each $z\geq 0$, the first part of the assertion follows. 
Now we show \eqref{regularity of the integral}. It is clear that 
$$
z\longrightarrow \int_{z}^{z+t}r(0,s,x_i)ds
$$
is continuous. Since
$$
z\rightarrow \int_{z}^{z+t-u}F_i(u,r(u))(v)dv
$$
is continuous and for each $z$ 
$$
\left\lvert \int_{z}^{z+t-u}F_i(u,r(u))(v)dv\right\rvert\leq  \int_{0}^{+\infty}\mid F_i(u,r(u))(v)\mid dv
$$
with
$$
\int_{0}^{t}\int_{0}^{+\infty}\mid F_i(u,r(u))(v)\mid dvdu<+\infty,
$$
it follows from the dominated convergence theorem that the function
$$
z\rightarrow \int_{0}^{t}\int_{z}^{z+t-u}F_i(u,r(u))(v)dvdu
$$ 
is continuous. Consider any $z\geq0$ and a sequence $z_n\rightarrow z$. 
Then for $\varepsilon>0$ and sufficiently large $n$ holds
\begin{align*}
&E\int_{0}^{t}\left\lvert \int_{z_n}^{z_n+t-u}G_i(u,r(u-))(v)dv-\int_{z}^{z+t-u}G_i(u,r(u-))(v)dv\right\rvert^2 du\\[1ex]
&\leq 4E\int_{0}^{t}\left(\int_{z-\varepsilon}^{z+t+\varepsilon}\mid G_i(u,r(u-))(v)\mid dv\right)^2du\\[1ex]
&\leq 4(2\varepsilon+t)E\int_{0}^{t}\int_{0}^{+\infty}\mid G_i(u,r(u-))(v)\mid^2 dv du<+\infty.
\end{align*}
Since the function 
$$
z\rightarrow \int_{z}^{z+t-u}G_i(u,r(u-))(v)dv
$$
is continuous it follows that
$$
E\int_{0}^{t}\left\lvert \int_{z_n}^{z_n+t-u}G_i(u,r(u-))(v)dv-\int_{z}^{z+t-u}G_i(u,r(u-))(v)dv\right\rvert^2 du\underset{n}{\longrightarrow} 0.
$$
That condition and \eqref{second moment of the measure} imply that 
$$
E\left(\int_{0}^{t}\int_{z_n}^{z_n+t-u}G_i(u,r(u-))(v)dvdZ(u)-\int_{0}^{t}\int_{z}^{z+t-u}G_i(u,r(u-))(v)dvdZ(u)\right)^2\longrightarrow 0,
$$
and thus we can find a subsequence $z_{n_k}$ such that 
$$
\int_{0}^{t}\int_{z_{n_k}}^{z_{n_k}+t-u}G_i(u,r(u-))(v)dvdZ(u)\underset{k}{\longrightarrow}\int_{0}^{t}\int_{z}^{z+t-u}G_i(u,r(u-))(v)dvdZ(u)
$$
almost surely. This leads to \eqref{regularity of the integral}.
\hfill$\square$

\begin{prop}\label{prop positivity in L2}
 Let $r(t), t\geq 0$ be a c\`adl\`ag nonnegative process taking values in $\mathbb{L}^{2,\gamma}_n$. Then
 for each $x_i\in I$ we have
 \begin{gather*}
  \int_{0}^{t}r(s,z,x_i)ds\geq 0,
 \end{gather*}
for each $t\geq0$ and almost all $z\geq 0$.
\end{prop}
{\bf Proof:} Since the process $r(t), t\geq0$ is c\`adl\`ag, it follows that for each $x_i\in I$ the function $t\rightarrow \|r(t,\cdot,x_i)\|_{L^{2,\gamma}}$ is bounded on bounded intervals. Thus for any $0\leq a<b\leq+\infty$ we have
\begin{align*}
 &\int_{0}^{t}\int_{a}^{b}r(s,z,x_i)dzds\leq\int_{0}^{t}\int_{0}^{+\infty}r(s,z,x_i)e^{\frac{\gamma}{2}z}e^{-\frac{\gamma}{2}z}dzds\\[1ex]
 &\leq \int_{0}^{t}\left(\int_{0}^{+\infty}r^2(s,z,x_i)e^{\gamma z}dz\right)^{\frac{1}{2}}\left(\int_{0}^{+\infty}e^{-\gamma z}dz\right)^{\frac{1}{2}}ds\leq\frac{1}{\sqrt{\gamma}}\int_{0}^{t}\|r(s,\cdot,x_i)\|_{L^{2,\gamma}}ds<+\infty.
\end{align*}
Thus the Fubini theorem yields
\begin{gather*}
0\leq\int_{0}^{t}\int_{a}^{b}r(s,z,x_i)dzds=\int_{a}^{b}\int_{0}^{t}r(s,z,x_i)dsdz. 
\end{gather*}
Since $a,b$ are arbitrary, the assertion follows. \hfill $\square$

\vskip2mm
Now we can easily prove  Proposition \ref{prop point monotonicity}.
\vskip2mm
\noindent
{\bf Proof of Proposition \ref{prop point monotonicity}:} From monotonicity of the solution and Proposition \ref{prop positivity in L2} follows that for each $t\geq0$ and $x_i\in I$
$$
\int_{0}^{t}r(s,z,x_i)ds\geq\int_{0}^{t}r(s,z,x_{i+1})ds \qquad \text{for almost all} \ z\geq 0.
$$
In view of Proposition \ref{prop jednoznacznosc calki w L2} the inequality above holds for each $z\geq0$. As a consequence, for each $z\geq0$ we have
$$
r(t,z,x_i)\geq r(t,z,x_{i+1}), \qquad \text{for almost all} \ t\geq0.
$$
\hfill $\square$

\vskip2mm
\noindent
{\bf Proof of Proposition \ref{prop monotonicity of forward rates}:} Since $r$ is a solution of the HJMM equation, it follows that the process $\hat{P}(t,T,x_i)$ is a local martingale for each $T>0, x_i\in I$. From the positivity of $r$ follows
$$
0\leq \hat{P}(t,T,x_i)=e^{-\int_{0}^{t}r(s,0,1)ds}\mathbf{1}_{\{L_t\leq x_i\}} e^{-\int_{0}^{T-t}r(t,u,x_i)du}\leq 1,
$$
and thus $\hat{P}(t,T,x_i)$ is a martingale. From the identity
$$
\hat{P}(T,T,x_i)=e^{-\int_{0}^{T}r(s,0,1)ds}\mathbf{1}_{\{L_t\leq x_i\}},
$$
follows that
$$
\hat{P}(t,T,x_i)=E(e^{-\int_{0}^{T}r(s,0,1)ds}\mathbf{1}_{\{L_t\leq x_i\}}\mid\mathcal{F}_t),
$$
and as a consequence one obtains
$$
P(t,T,x_i)=e^{\int_{0}^{t}r(s,0,1)ds}E(e^{-\int_{0}^{T}r(s,0,1)ds}\mathbf{1}_{\{L_t\leq x_i\}}\mid\mathcal{F}_t)=E(e^{-\int_{t}^{T}r(s,0,1)ds}\mathbf{1}_{\{L_t\leq x_i\}}\mid\mathcal{F}_t).
$$
Thus monotonicity of $P(t,T,x_i)$ in $x_i$ follows. Now assume to the contrary that 
$$
r(t,0,x_i)<r(t,0,x_{i+1}) \quad \text{for some} \ t\geq0, L_t\leq x_i, i\in\{1,2,...,n-1\}.
$$
Since $r(t)\in\mathbb{H}^{1,\gamma}_n$ the function $z\rightarrow r(t,z,x_i)$ is continuous and thus 
$$
r(t,u,x_i)<r(t,u,x_{i+1}), \qquad  u\in(0,\varepsilon), 
$$
for some $\varepsilon>0$. This implies
$$
P(t,t+\varepsilon,x_i)=e^{-\int_{0}^{\varepsilon}r(t,u,x_i)du}>e^{-\int_{0}^{\varepsilon}r(t,u,x_{i+1})du}=P(t,t+\varepsilon,x_{i+1})
$$
which is a contradiction to monotonicity of $P(t,T,x_i)$ in $x_i$ proved above. \hfill $\square$

\subsection{Proof of Proposition \ref{prop o konkretnych warunkach na positivity} and calculations for the Example \ref{przyklad}}
\label{section ostatni}
{\bf Proof of Proposition \ref{prop o konkretnych warunkach na positivity}:} $(A)$ $(a)$ It follows from the condition (P2)  and positivity of
$g_i$ that
\begin{gather*}
 u\geq -\frac{r_i}{g_i(t,z,l,r)}, \quad t,z,r\geq0, \ l\in I,  \ u\in\text{supp}\{\nu\}, \ i=1,2,...,n.
\end{gather*}
It follows from (P1) that
\begin{gather*}
 u\geq -\frac{r_i}{g_i(t,z,l,r)-g_i(t,z,l,\mathbf{1}_i(r))}.
 \end{gather*}
Passing with $r\downarrow 0$ and taking supremum over $t,z,l,r$ yields \eqref{war konieczny positivity - g rozniczkowalna}.\\
\noindent $(b)$ It follows from \eqref{war konieczny positivity - g rozniczkowalna} that
\begin{gather*}
u\geq -\frac{1}{g^{\prime}_{r_i}(t,z,x_i,l,\mathbf{1}_i(r))}, \qquad t,z\geq 0, \ r\geq0, \ l\in I, \ u\in\text{supp}\{\nu\}, i=1,2,...,n.
\end{gather*}
Using \eqref{war dostateczny positivity - g rozniczkowalna} yields 
\begin{gather*}
u\geq -\frac{1}{g^{\prime}_{r_i}(t,z,x_i,l,\mathbf{1}_i(r))}\geq-\frac{r_i}{g_i(t,z,l,r)}
\end{gather*}
which is (P2).\\
\noindent
$(B)$ $(a)$ In view of (M1) condition (M2) is equivalent to
\begin{gather*}
\Big(\frac{g_{i+1}(t,z,l,r_1,...,r_{i},r_{i+1},...,r_n)-g_{i+1}(t,z,l,r_1,...,r_{i+1},r_{i+1},...,r_n)}{r_i-r_{i+1}}\\[1ex]
-\frac{g_i(t,z,l,r_1,...,r_{i},r_{i+1},...,r_n)-g_i(t,z,l,r_1,...,r_{i+1},r_{i+1},...,r_n)}{r_i-r_{i+1}}\Big)u\leq 1,
\end{gather*}
for $r\in\mathbb{R}^n, r_1\geq r_2\geq...\geq r_n, \ t,z\geq0, \ l\in I, \ u\in\emph{supp}\{\nu\}$, $i=1,2,...,n-1$.
Passing to the limit $r_i\downarrow r_{i+1}$ yields \eqref{pierwszy war na roznice pochodnych}. \eqref{drugi war na roznice pochodnych} 
can be obtained in a similar way.\\
\noindent
$(b)$  
The concavity of $g_{i+1}(t,z,l,r)-g_{i}(t,z,l,r)$ in $r_i$ and (M1) imply
\begin{gather}\nonumber
\frac{g_{i+1}(t,z,l,r_1,...,r_{i},r_{i+1},...,r_n)-g_{i}(t,z,l,r_1,...,r_{i},r_{i+1},...,r_n)}{r_i-r_{i+1}}\leq\\[1ex]\label{wkleslosc roznicy}
\frac{d}{dr_i}\left[g_{i+1}(t,z,l,r_1,...,r_{i+1},r_{i+1},...,r_n)-g_{i+1}(t,z,l,r_1,...,r_{i+1},r_{i+1},...,r_n)\right],
\end{gather}
while convexity gives
\begin{gather}\nonumber
\frac{g_{i+1}(t,z,l,r_1,...,r_{i},r_{i+1},...,r_n)-g_{i}(t,z,l,r_1,...,r_{i},r_{i+1},...,r_n)}{r_i-r_{i+1}}\geq\\[1ex]\label{wypuklosc roznicy}
\frac{d}{dr_i}\left[g_{i+1}(t,z,l,r_1,...,r_{i+1},r_{i+1},...,r_n)-g_{i+1}(t,z,l,r_1,...,r_{i+1},r_{i+1},...,r_n)\right].
\end{gather}
If $\text{supp}\{\nu\}\subseteq(0,+\infty)$, then multiplying both sides of \eqref{wkleslosc roznicy} with $u>0$ and using \eqref{pierwszy war na roznice pochodnych} yields (M2). Thus $(i)$ implies (M2). If $\text{supp}\{\nu\}\subseteq(-\infty,0)$, then multiplying both sides of \eqref{wypuklosc roznicy} with $u<0$ and using \eqref{pierwszy war na roznice pochodnych} yields (M2) 
and shows sufficiency of $(ii)$.
If $(iii)$ holds, then (M2) is satisfied for each $u<0$. For $u>0$ we can argue as in $(1)$.
Similarly $(iv)$ implies (M2) for $u>0$ while for $u<0$ we use the same argument as in $(ii)$. \hfill $\square$

\vskip2mm

\noindent
{\bf Calculations for the Example \ref{przyklad}}
 $(a)$ It is clear that \eqref{P2} implies (P1). We have
$$
\frac{d}{dr_i}g_i(t,z,l,r)=f_1(t)f_2(z)f_3(l)h_1(r_1)...h_{i-1}(r_{i-1})\Big[h_i^\prime(r_i)h(r_i)+h_i(r_i)h^\prime(r_i)\Big]h_{i+1}(r_{i+1})...h_n(r_n),
$$
which gives
$$
\frac{d}{dr_i}g_i(t,z,l,r)=f_1(t)f_2(z)f_3(l)h_1(r_1)...h_{i-1}(r_{i-1})\Big[h_i(0)h^\prime(0)\Big]h_{i+1}(r_{i+1})...h_n(r_n)\geq0,
$$
for $r_i=0$. Thus \eqref{P4} implies necessary condition \eqref{war konieczny positivity - g rozniczkowalna}. Further \eqref{P2} and \eqref{P3} imply inequality
$$
h_i(r_i)h(r_i)\leq h_i(0)h^\prime(0) r_i, \qquad \ r_i\geq0,
$$
which is exactly \eqref{war dostateczny positivity - g rozniczkowalna}. Hence it follows from Proposition \ref{prop o konkretnych warunkach na positivity} $(A)$ that (P2) is satisfied.
\vskip2mm
\noindent
$(b)$ It is clear that (M1) holds. If $r_i=r_{i+1}$, the following holds
$$
\frac{d}{dr_i}\Big(g_{i+1}(t,z,l,r)-g_i(t,z,l,r)\Big)=-f_1(t)f_2(z)f_3(l)h_1(r_1)...h_n(r_n)h^\prime(r_i),
$$
and thus condition \eqref{pierwszy war na roznice pochodnych} is implied by \eqref{P5} and \eqref{P6}. Moreover, we have
$$
g_{i+1}(t,z,l,r)-g_i(t,z,l,r)=f_1(t)f_2(z)f_3(l)h_1(r_1)...h_n(r_n)[h(r_{i+1})-h(r_i)], \quad r_1\geq r_2\geq... r_n,
$$
and thus for convexity in $r_i$ we need to examine convexity of the function
$$
r_i\rightarrow h_i(r_i)[h(r_{i+1})-h(r_i)], \quad r_i\geq r_{i+1},
$$
In view of \eqref{P1}, \eqref{P2} and \eqref{P5} we have
$$
\frac{d^2}{dr_i^2}\Big[h_i(r_i)[h(r_{i+1})-h(r_i)] \Big]=h_i^{\prime\prime}(r_i)[h(r_{i+1})-h(r_i)]-2h^\prime_i(r_i)h^\prime(r_i)-h_i(r_i)h^{\prime\prime}(r_i)\geq0,
$$
and convexity follows. It is also clear that $g_{i+1}(t,z,l,r)\leq g_i(t,z,l,r)$. Thus, in view of Proposition \ref{prop o konkretnych warunkach na positivity} (B), condition (M2) is satisfied.\hfill$\square$

\vskip2mm
\noindent
{\bf Acknowledgements}\\
The author would like to thank A. Rusinek, T. Schmidt, S. Tappe and J. Zabczyk for inspiring discussions and helpful suggestions. The paper was supported by The Polish MNiSW Grant NN201419039.

\end{document}